\documentclass[twocolumn,showpacs,a4paper]{revtex4}
\usepackage{graphicx}
\usepackage{amssymb}
\usepackage{color}
\usepackage{ulem}

\begin{document}

\title{Entanglement of superconducting qubits via microwave fields: classical and quantum
regimes}

\author{Jian Li}

\author{K. Chalapat}\altaffiliation{Also at the NanoScience Center and Department of Physics,
University of Jyv\"askyl\"a, P.O.~Box 35 (YFL), FIN-40014 University
of Jyv\"askyl\"a, Finland}

\author{G. S. Paraoanu}\email{paraoanu@cc.hut.fi}

\affiliation{Low Temperature Laboratory, Helsinki University of Technology,
P.O. Box 5100, FIN-02015 TKK, Finland}

\begin{abstract}

We study analytically and numerically the problem of two qubits with fixed coupling irradiated with quantum or classical fields.
In the classical case, we derive an effective Hamiltonian and describe its entangling properties. We identify a coupling/decoupling switching
protocol
and we construct composite pulse sequences leading to a CNOT gate. In the
quantum case, we show that qubit-qubit-photon multiparticle entanglement and maximally entangled two-qubit states can be obtained
by driving the system at very low powers (one quanta of excitation). Our results can be applied to a variety of systems of two
superconducting qubits coupled to resonators.

\end{abstract}


\maketitle

\section{Introduction}
\label{intro}
In recent years, there has been a consistent experimental progress in the quantum-coherent
manipulation of superconducting circuits based on the Josephson effect. Single-qubit
operations have been demonstrated by now in charge \cite{charge}, flux \cite{flux},
charge-flux \cite{chargeflux}, and phase \cite{phase} qubits.
Some of these experiments \cite{chargeflux} made it clear that in order to achieve long decoherence times the qubits
have to be operated at the so-called optimal point, where the first order noise induced by fluctuations of external control parameters (gate voltages, magnetic fluxes) cancels. The notable exception from this rule is the phase qubit, which by construction does not have an optimal point; this is compensated by engineering a peculiar bias circuit \cite{phase}.

The first-generation experiments with fixed coupling \cite{first1,first2} did manage to achieve two-qubit gates but the qubits were not
operated at the optimal points, with a corresponding loss in fidelity.
Later it was noticed that flux qubits can be operated at the optimal point if the coupling is realized
through a dc-SQUID modulated at the sum and difference of the qubits' resonance frequencies \cite{bertet}.
If the dc-SQUID is replaced by a third flux qubit, largely detuned from the frequencies of the two qubits, the
coupling can be realized through the quantum inductance of the additional qubit \cite{niskanen}.

With the advent of circuit QED architectures \cite{yale1, yale2}, several qubits can be placed in the gap between the
signal line and the ground of a coplanar waveguide resonator. In the dispersive qubit-resonator coupling regime,
the qubit-qubit coupling are realized by virtually exciting the resonator. Furthermore, due to the structured, cleaner
electromagnetic environment around the qubits, relatively long decoherence time can be achieved.

Finally, since any additional coupling elements tend to introduce extra decoherence in the system, one can place a
further restriction and ask the question: is it possible to devise schemes in which the qubits are operated at the optimal
points and there are no active additional elements for coupling? The first such proposal was inspired from NMR: the so-called
FLICFORQ protocol \cite{Rigetti}, with coupling realized through the dressed states of each qubit under on-resonance microwave fields.
The protocol was extended to more general detuned driving fields \cite{Ashhab}, but although theoretically sound, it
has not been realized yet experimentally. In the FLICFORQ protocol, the fixed coupling is much smaller than the qubit-qubit detuning;
even under the enhancement
the effective coupling strength is up to only one eighth of the bare coupling strength \cite{yale2, Rigetti}.  A small coupling strength is
desirable for single-qubit
operations, but limit the two-qubit gate speed.

Therefore, one very interesting question emerges: is it possible to develop a protocol
which retains the advantages of FLICFORQ and has also the capability of performing fast two-qubit gates?
In this work we will analyze such a protocol, in which the two qubits are relatively strong coupled in the absence of driving fields; moreover,
we show explicitly how to construct microwave pulses to switch the coupling off.

The outline of this paper is as follows. In Sec. \ref{model} we study a specific circuit consisting of two qubits coupled to a
single cavity mode (a coplanar waveguide resonator). A general fixed qubit-qubit coupling Hamiltonian is derived.
We then analyze the case in which the qubits are manipulated using classical fields: in Secs. \ref{switchable} and
\ref{RWA}, by deriving the effective Hamiltonian of the system in the rotating reference frame, we show that for large values of the
driving field amplitudes, the coupling is switchable. We also find numerically the on/off coupling ratio of the switchable coupling,
as well as the validity of our rotating wave approximation (RWA). The implementations of a single-qubit gate and a CNOT gate are
demonstrated in Sec. \ref{quantum_gates}. We show that both gates can be realized with high speed and high fidelity.
In the quantum case, in which the qubits interact with a single quanta of radiation,
we propose a quantum nondemolition (QND) method to entangle the qubits
based on measuring the presence of an excitation in the resonant cavity to entangle the qubits in Sec. \ref{QND}, and finally we conclude our
work in Sec. \ref{conclude}.

\section{Model Hamiltonians}
\label{model}

\begin{figure}[htb]
\includegraphics[width=8cm]{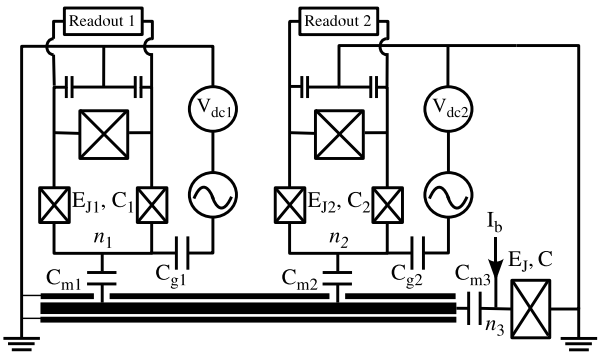}
\caption{Schematic circuit of our model system.}
\label{fig_system}
\end{figure}

Although many of the results derived in this paper are rather general (once the Hamiltonian is put into any of the forms used in this paper),
it is useful to start by analyzing a concrete superconducting quantum circuit, which will serve as our workhorse.
Consider a system of two coupled charge qubits, irradiated with monochromatic off-resonance microwave fields.
The circuit is shown in Fig. \ref{fig_system}. Two split single Cooper pair box (SCB) qubits are capacitively coupled to the
center conductor of a coplanar waveguide resonator. Each box has its own read-out circuitry (such as a large current-biased Josephson junction,
as in the case of charge-flux qubits).

To ensure first-order insensitivity to charge fluctuations, we voltage-bias the gates of the SCBs at the optimal points
\begin{equation}
C_{g1}V_{\mathrm{dc}1} = C_{g2}V_{\mathrm{dc}2} = -e . \label{eq_charge_degeneracy}
\end{equation}
When two microwave fields with the same angular frequency $\omega_d$ and different phases $\phi_{1,2}$ are applied, the gate
voltages of the SCBs have AC components
\begin{equation}
V_{\mathrm{ac}1,2} = V_{\mathrm{\mu w}1,2}(t)\cos(\omega_d t + \phi_{1,2}) . \label{eq_ac_voltage}
\end{equation}

Written in the eigenbasis of each SCB $|\uparrow\rangle = (|0\rangle + |1\rangle) / \sqrt{2}$ and
$|\downarrow\rangle = (|0\rangle - |1\rangle) / \sqrt{2}$, $|0\rangle$ and $|1\rangle$ denoting two lowest charge states, the
Hamiltonian has the following form (see Appendix \ref{resonator_coupling})
\begin{equation}
H_\mathrm{SCB} = \sum_{j=1,2}\left[ - \frac{E_{Jj}}{2}\sigma_j^z + E_{Cj}w_j(t)\cos(\omega_dt + \phi_j)\sigma_j^x \right] , \label{eq_h_scb}
\end{equation}
where $E_{Jj}$ and $E_{Cj}$ indicate the standard Josephson and charging energies, and $w_j(t) \equiv -C_{gj}V_{\mathrm{\mu w}j}(t) / 2e$.

A large current-biased Josephson junction (CBJJ) is also coupled
to the end of the resonator through a capacitor; the
bias current $I_b$ is such that, during many of the qubit operations described in this paper, only two bound states
(of energy difference $\omega_{10}$
close to that of the resonator) are relevant. This junction will serve as a detector of the state of the resonator, as we will see in
Sec. \ref{QND}.
The Hamiltonian of the CBJJ assumes the form of a two-level system,
\begin{equation}
H_\mathrm{CBJJ} = -\frac{\omega_{10}}{2}\sigma_3^z ,
\end{equation}
where $\omega_{10}$ denotes the transition frequency between the two lowest levels in the tilted cosine potential.

To simplify the equations, from now on our units will be such that  $\hbar = 1$.

Depending on the length of the stripline, the qubit-qubit and qubit-CBJJ couplings must be treated in different ways. For a relatively long
resonator, the couplings are mediated by excitations in the resonator. As derived in Appendix \ref{resonator_coupling}, the total Hamiltonian
in this case is
\begin{eqnarray}
H_\mathrm{tot1} &=& H_\mathrm{SCB} + H_\mathrm{CBJJ} + \omega_r(a^\dag a + 1/2) \nonumber \\
&& + i \sum_{j=1,2} g_j \left[ \sigma_j^x + 2w_j(t)\cos(\omega_dt + \phi_j) \right] (a^\dag - a) \nonumber \\
&& + \kappa (\sigma_3^+ - \sigma_3^-) (a^\dag - a) , \label{eq_ham_long_res1}
\end{eqnarray}
with $\omega_r$ the resonance frequency of the resonator, $a^\dag$ ($a$) the photon creation (annihilation) operator,
$g_j$ ($\kappa$) the qubit-resonator (CBJJ-resonator) coupling strength, and $\sigma_3^+$ ($\sigma_3^-$) the CBJJ raising (lowering) operator. By performing a RWA to neglect fast oscillating terms, we have
\begin{eqnarray}
H_\mathrm{tot1} &\approx & H_\mathrm{SCB} + H_\mathrm{CBJJ} + \omega_r(a^\dag a + 1/2) \nonumber \\
&& + i\sum_{j=1,2} g_j (\sigma_j^- a^\dag - \sigma_j^+ a) \nonumber \\
&& - \kappa (\sigma_3^+ a + \sigma_3^- a^\dag) , \label{eq_ham_long_res2}
\end{eqnarray}
where $\sigma_j^+ = |\downarrow\rangle \langle\uparrow|$ and $\sigma_j^- = |\uparrow\rangle \langle\downarrow|$ denote
qubit raising and lowering operators, respectively. Here we have neglected the terms describing interactions between the qubit driving
fields and the resonator (in the second line of Eq. (\ref{eq_ham_long_res1})).
This is justified by the fact that Rabi frequencies satisfy $\Omega_j(t) \equiv E_{Cj}w_j(t) \ll E_{Jj}$, and the SCBs are working in the charge
regime $E_{Cj}\gg E_{Jj}$, therefore $w_j(t)$
should be much smaller than unity.

\paragraph{Long resonator limit \newline}

In the long resonator limit (the resonator length of the same order as the wavelength corresponding to
an excitation with energy of the order of qubit energy), a fixed linear qubit-qubit coupling can be achieved in
the dispersive regime, $g_j\ll \Delta_j \equiv \omega_r - E_{Jj}$.
By performing a generalized Schrieffer-Wolff transformation (see Appendix \ref{eff_resonator_coupling})
\begin{eqnarray}
e^A &=& \exp\left[ -i\sum_{j=1,2} \frac{g_j}{\Delta_j} ( a^\dag\sigma_j^- + a\sigma_j^+ ) \right. \nonumber \\
&& \left. \ \ \ \ \ \ \ -i\sum_{j=1,2} \frac{g_j\Omega_j(t)\cos(\omega_d t + \phi_j)}{\Delta_j\omega_r} \sigma_j^z (a^\dag + a) \right] \nonumber
\end{eqnarray}
on the Hamiltonian (\ref{eq_ham_long_res2}), we obtain
\begin{eqnarray}
\widetilde{H}_\mathrm{tot1} &\approx& \sum_{j=1,2}\left[ -\frac{E_{Jj}}{2}\sigma_j^z + \Omega_j\cos(\omega_dt + \phi_j)\sigma_j^x \right] \nonumber \\
&& - \frac{g_1g_2(\Delta_1 + \Delta_2)}{2\Delta_1 \Delta_2}(\sigma_1^+\sigma_2^- + \sigma_1^-\sigma_2^+) . \label{eq_h_qb-res}
\end{eqnarray}
Here we assume that the CBJJ is biased far off resonance with the resonator such that it is effectively decoupled from it.

\paragraph{Short resonator limit \newline}

For a very short resonator (the resonator length much smaller than the wavelength corresponding to an excitation with energy of the order of qubit energy), the center conductor can be considered as a small metallic island. The two qubits and the CBJJ are capacitively coupled to this island.
In Appendix \ref{capacitor_coupling} we show that the total Hamiltonian has the form
\begin{eqnarray}
H_\mathrm{tot2} &=& H_\mathrm{SCB} + H_\mathrm{CBJJ}  + \frac{E_{12}}{4}\sigma_1^x\sigma_2^x - \sum_{j=1,2} \frac{\gamma_j}{2}\sigma_j^x\sigma_3^y \nonumber \\
&& + \frac{E_{12}}{2} [ w_2(t)\cos(\omega_dt + \phi_2)\sigma_1^x \nonumber \\
&& \ \ \ \ \ \ \ \ \ + w_1(t)\cos(\omega_dt + \phi_1)\sigma_2^x ] \nonumber \\
&& - \sum_{j=1,2}\gamma_j w_j(j)\cos(\omega_d t + \phi_j)\sigma_3^y , \label{eq_ham_short_res1}
\end{eqnarray}
with $E_{12} / 4$ and $\gamma_j / 2$ the qubit-qubit and qubit-CBJJ coupling strengths, respectively.
Due to small coupling $E_{12}\ll E_{C1,2}$, the {\it crosstalk} \cite{Ashhab} terms in the second and third lines
of Eq. (\ref{eq_ham_short_res1}) are negligible. We can also drop the interactions between the qubit driving fields and the CBJJ,
described by the last line in Eq. (\ref{eq_ham_short_res1}), because $w_j$ is small. Therefore the total Hamiltonian is approximately
\begin{equation}
H_\mathrm{tot2} \approx H_\mathrm{SCB} + H_\mathrm{CBJJ}  + \frac{E_{12}}{4}\sigma_1^x\sigma_2^x - \sum_{j=1,2}\frac{\gamma_j}{2}\sigma_j^x\sigma_3^y . \label{eq_ham_short_res2}
\end{equation}
If the coupling capacitances $C_{m1,2}\gg C_{m3}$, the direct coupling is dominating
\begin{eqnarray}
H_\mathrm{tot2} &\approx& \sum_{j=1,2}\left[ -\frac{E_{Jj}}{2}\sigma_j^z + \Omega_j\cos(\omega_dt + \phi_j)\sigma_j^x \right] \nonumber \\
&& + \frac{E_{12}}{4} \sigma_1^x\sigma_2^x . \label{eq_h_direct_coupling}
\end{eqnarray}
For $C_{m1,2}\ll C_{m3}$, the two qubits can be coupled through virtual excitation of the CBJJ. By considering dispersive
coupling $\gamma_j\ll \Delta_j' \equiv \omega_{10} - E_{Jj}$, the qubit-CBJJ couplings are eliminated by another
Schrieffer-Wolff transformation similar to the one in long resonator case, and we obtain
\begin{eqnarray}
\widetilde{H}_\mathrm{tot2} &\approx& \sum_{j=1,2}\left[ -\frac{E_{Jj}}{2}\sigma_j^z + \Omega_j\cos(\omega_dt + \phi_j)\sigma_j^x \right] \nonumber \\
&& - \frac{\gamma_1\gamma_2(\Delta_1' + \Delta_2')}{4\Delta_1' \Delta_2'}(\sigma_1^+\sigma_2^- + \sigma_1^-\sigma_2^+) . \label{eq_h_qb-cbjj}
\end{eqnarray}

The Hamiltonians (\ref{eq_h_qb-res}) (\ref{eq_h_direct_coupling}) and (\ref{eq_h_qb-cbjj}) are similar, therefore
we will use a generic time-dependent Hamiltonian
\begin{equation}
H(t) = \sum_{j = 1,2} \left[ - \frac{\omega_j^L}{2}\sigma_j^z + \Omega_j\cos(\omega_dt + \phi_j)\sigma_j^x \right] + \omega^{xx} \sigma_1^x\sigma_2^x \label{eq_time_dep_ham}
\end{equation}
to replace all of them. $\omega_j^L$ and $\omega^{xx}$ denote the Larmor frequency of qubit-$j$ and the qubit-qubit coupling strength, respectively. As discussed in Sec. \ref{intro}, we are interested in the regime $\omega^{xx}\approx |\omega_1^L - \omega_2^L| \ll \omega_{1,2}^L$.

\section{Switchable coupling mechanism}
\label{switchable}

Now, we start to derive an effective Hamiltonian with tunable coupling terms. In order to eliminate the explicit time dependence, we bring the Hamiltonian (\ref{eq_time_dep_ham}) into the rotating reference frame by transforming it with the operator
\begin{equation}
S_1(t) = \exp\left[ \frac{i\omega_d t}{2}(\sigma_1^z + \sigma_2^z)
\right] . \label{eq_unit_oper1}
\end{equation}
By performing a RWA to neglect oscillating terms with frequency $2\omega_d$, we get a time-independent effective Hamiltonian
\begin{equation}
H_{\mathrm{eff}} = H_{12} + \sum_{j=1,2} H_j, \label{eq_eff_ham0}
\end{equation}
where
\begin{equation}
H_j = \frac{\delta_j}{2}\sigma_j^z +
\frac{\Omega_j}{2}(\cos\phi_j\sigma_j^x - \sin\phi_j\sigma_j^y)
\label{eq_eff_hamj}
\end{equation}
is the Hamiltonian for qubit-$j$, and
\begin{equation}
H_{12} = \frac{\omega^{xx}}{2}(\sigma_1^x\sigma_2^x + \sigma_1^y\sigma_2^y)
\label{eq_eff_ham12}
\end{equation}
describes the interaction between the two qubits. $\delta_j \equiv \omega_d - \omega_j^L$ denotes the detuning between the driving frequency and the corresponding qubit Larmor frequency.

By diagonalizing the single-qubit Hamiltonians given by Eq.
(\ref{eq_eff_hamj}), we obtain the eigenenergies and the corresponding eigenstates of each single qubit:
\begin{eqnarray}
&& E_+^{(j)} = \widetilde{\omega}_j / 2, \ \ \ \ \ \ \ \ \ \ \ \ \ \ \ \ \ \ \ \ \ E_-^{(j)} = -\widetilde{\omega}_j / 2, \nonumber \\
&& |+\rangle^{(j)} = \cos\frac{\theta_j}{2}e^{i\phi_j/2}|\uparrow\rangle
+ \sin\frac{\theta_j}{2}e^{-i\phi_j/2}|\downarrow\rangle , \label{eq_eigenstate1} \\
&& |-\rangle^{(j)} = -\sin\frac{\theta_j}{2}e^{i\phi_j/2}|\uparrow\rangle
+ \cos\frac{\theta_j}{2}e^{-i\phi_j/2}|\downarrow\rangle , \ \ \ \ \label{eq_eigenstate2}
\end{eqnarray}
where
\begin{equation}
\widetilde{\omega}_j = \sqrt{\delta_j^2 + \Omega_j^2} , \ \ \sin\theta_j = \frac{\Omega_j}{\widetilde{\omega}_j}, \ \ \cos\theta_j = \frac{\delta_j}{\widetilde{\omega}_j} . \label{eq_eigenfreqs}
\end{equation}

We then project the time-independent effective Hamiltonian (\ref{eq_eff_ham0}) onto the new basis of product states $|+ +\rangle$, $|+ -\rangle$, $|- +\rangle$, $|- -\rangle$, and get
\begin{eqnarray}
H_{\mathrm{eff}}' &=& \frac{\widetilde{\omega}_1}{2}\sigma_z^{(1)} +
\frac{\widetilde{\omega}_2}{2}\sigma_z^{(2)} \nonumber \\
&& + \frac{\omega^{xx}}{2}\cos\phi \left\{ \left[ \sin\theta_1\sigma_z^{(1)} + \cos\theta_1\sigma_x^{(1)} \right] \right. \nonumber \\
&& \ \ \ \ \ \ \left. \times \left[ \sin\theta_2\sigma_z^{(2)} + \cos\theta_2\sigma_x^{(2)} \right] + \sigma_y^{(1)}\sigma_y^{(2)} \right\} \nonumber \\
&& + \frac{\omega^{xx}}{2}\sin\phi \left\{ \sigma_y^{(1)} \left[ \sin\theta_2\sigma_z^{(2)} + \cos\theta_2\sigma_x^{(2)} \right] \right. \nonumber \\
&& \ \ \ \ \ \ - \left. \left[ \sin\theta_1\sigma_z^{(1)} + \cos\theta_1\sigma_x^{(1)} \right] \sigma_y^{(2)} \right\} , \label{eq_eff_ham_pri}
\end{eqnarray}
with the Pauli matrices $\sigma_s^{(1)}$ and $\sigma_s^{(2)}$ ($s = x,y,z$) in the new basis,
and the phase difference $\phi \equiv \phi_1 - \phi_2$.

Next we transform $H_{\mathrm{eff}}'$ into a new rotating frame with
\begin{equation}
S_2(t) = \exp\left\{ \frac{-it}{2}\left[ \widetilde{\omega}_1\sigma_z^{(1)} + \widetilde{\omega}_2\sigma_z^{(2)} \right] \right\}. \label{eq_unit_oper2}
\end{equation}
By assuming that $\widetilde{\omega}_{1,2}$ are much larger than the coupling strength $\omega^{xx}$, we may perform a second RWA to neglect oscillating terms with frequencies $\widetilde{\omega}_{1,2}$ and $\widetilde{\omega}_1+\widetilde{\omega}_2$. The resulting Hamiltonian is
\begin{widetext}
\begin{eqnarray}
H_\mathrm{eff}'' &=& \frac{\omega^{xx}}{4}\cos\phi(1+\cos\theta_1\cos\theta_2)\left\{ \cos(\delta\widetilde{\omega} t) \left[ \sigma_x^{(1)}\sigma_x^{(2)} + \sigma_y^{(1)}\sigma_y^{(2)} \right] - \sin(\delta\widetilde{\omega} t) \left[ \sigma_y^{(1)} \sigma_x^{(2)} - \sigma_x^{(1)}\sigma_y^{(2)}\right] \right\} \nonumber \\
&& + \frac{\omega^{xx}}{4}\sin\phi(\cos\theta_1 + \cos\theta_2) \left\{ \sin(\delta\widetilde{\omega} t) \left[ \sigma_x^{(1)}\sigma_x^{(2)} + \sigma_y^{(1)}\sigma_y^{(2)} \right] + \cos(\delta\widetilde{\omega} t) \left[ \sigma_y^{(1)} \sigma_x^{(2)} - \sigma_x^{(1)}\sigma_y^{(2)}\right] \right\} \nonumber \\
&& + \frac{\omega^{xx}}{2}\cos\phi\sin\theta_1\sin\theta_2\sigma_z^{(1)}\sigma_z^{(2)} , \label{eq_eff_ham_double_pri1}
\end{eqnarray}
\end{widetext}
with $\delta\widetilde{\omega} \equiv \widetilde{\omega}_1 - \widetilde{\omega}_2$.
This effective Hamiltonian has oscillating terms. In the rest of this section, we discuss how to switch the effective coupling off in two limits with respect to the oscillation frequency $\delta\widetilde{\omega}$.

\subsection{High $\delta\widetilde{\omega}$}
\label{high_osc_freq}

The oscillating terms in (\ref{eq_eff_ham_double_pri1}) can be neglected if $|\delta\widetilde{\omega}| \gg \omega^{xx}$. By using the definition of $\sin\theta_j$ in (\ref{eq_eigenfreqs}), the remaining effective Hamiltonian is expressed as
\begin{equation}
H_\mathrm{eff}'' = \frac{\omega^{xx}}{2} \cos\phi \frac{\Omega_1 \Omega_2}{\sqrt{(\delta_1^2 + \Omega_1^2) (\delta_2^2 + \Omega_2^2)}} \sigma_z^{(1)}\sigma_z^{(2)} . \label{eq_eff_ham_double_pri2}
\end{equation}
A maximum coupling strength of about $\omega^{xx}/2$ can be achieved when the driving frequency is in-resonance with the qubit Larmor frequencies $\omega_d \approx \omega_{1,2}^L$, and large driving amplitudes $\Omega_{1,2}\gg |\delta_{1,2}|\approx \omega^{xx}$ are applied in the meantime.

To turn the coupling off, we can switch off either $\Omega_1$ or $\Omega_2$. Nevertheless, the two
conditions must be satisfied: $\widetilde{\omega}_{1,2}\gg \omega^{xx}$ and $|\delta\widetilde{\omega}|\gg \omega^{xx}$.
Without loss of generality, we assume $\Omega_2 = 0$. The condition $\widetilde{\omega}_{1,2} \gg \omega^{xx}$ is fulfilled when the driving frequency is
largely detuned from the qubit frequencies, $\delta_1\approx\delta_2 \gg \omega^{xx}$ (assuming $\omega_d > \omega_{1,2}^L$). The second
condition is fulfilled by driving the first qubit with a rather large amplitude $\Omega_1 \gg \sqrt{2\delta_1\omega^{xx}}$.

We notice that this decoupling mechanism can also be derived in the case of quantized driving fields: let us start with the usual Jaynes-Cummings
form \cite{Schleich} of the Hamiltonian
\begin{equation}
H = \omega_d a^\dag a -\sum_{j = 1,2}\frac{\omega_j^L}{2}\sigma_j^z + \omega^{xx}\sigma_1^x\sigma_2^x + g(\sigma_1^+a + \sigma_1^-a^\dag) , \label{eq_qubit_tls_resonator}
\end{equation}
and assume a dispersive coupling $g \ll \delta_1$. Then we perform a Schrieffer-Wolff transformation $U = \exp[g(\sigma_1^+a - \sigma_1^-a^\dag)/\delta_1]$ to eliminate the direct qubit-field coupling to leading order, and obtain
\begin{eqnarray}
U^\dag H U &\approx& \omega_d a^\dag a - \left[ \frac{\omega_1^L}{2} + \frac{g^2}{\delta_1} \left( a^\dag a + \frac{1}{2} \right) \right]\sigma_1^z \nonumber \\
&& - \frac{\omega_2^L}{2}\sigma_2^z + \omega^{xx}\sigma_1^x\sigma_2^x .
\end{eqnarray}
Due to the driving field applied on it, the Larmor frequency of qubit-1 is ac-Stark shifted by the
quantity $2 g^2 \langle a^\dag a\rangle / \delta_1$. If the ac-Stark shift is much larger than the qubit-qubit coupling strength $\omega^{xx}$, the
two qubits are effectively decoupled.


\subsection{Low $\delta\widetilde{\omega}$}
\label{low_osc_freq}

A more general situation is when $\delta\widetilde{\omega}$ is not so large, therefore the oscillating terms can not be neglected. Since we are considering a relatively small qubit-qubit detuning, it is not possible to eliminate $1 + \cos\theta_1 \cos\theta_2$, $\cos\theta_1 + \cos\theta_2$, $\sin\theta_1 \sin\theta_2$, and satisfy $\widetilde{\omega}_{1,2}\gg \omega^{xx}$ at the same time. However, there is still a way to switch the coupling off. As shown in Eq. (\ref{eq_eff_ham_double_pri1}), terms in the first and the third lines have a common factor $\cos\phi$. If the second line can be removed by setting $\delta_j$ and $\Omega_j$ to obtain
\begin{equation}
\cos\theta_1 + \cos\theta_2 = \frac{\delta_1}{\sqrt{\delta_1^2 + \Omega_1^2}} + \frac{\delta_2}{\sqrt{\delta_2^2 + \Omega_2^2}} = 0 , \label{eq_res_cond1}
\end{equation}
the rest of the Hamiltonian will be switchable by means of $\phi$. One solution of (\ref{eq_res_cond1}) is $\delta_1 = \delta_2 = 0$,
which is realized only when the qubits are on resonance. For off-resonance qubits, Eq. (\ref{eq_res_cond1}) leads to
\begin{equation}
\Omega_1 / \Omega_2 = -\delta_1 / \delta_2 .
\end{equation}

An extreme case is when $\delta\widetilde{\omega} = 0$. By defining $\Delta \equiv \omega_1^L - \omega_2^L$, $\delta_1 = \delta_2 - \Delta$, and using  Eq. (\ref{eq_eigenfreqs}) we obtain the expression of the resonance condition
\begin{equation}
\widetilde{\omega}_1 = \widetilde{\omega}_2 \ \Longrightarrow \ \Omega_1^2 = \Omega_2^2 + 2\delta_2\Delta - \Delta^2 , \label{eq_res_cond2}
\end{equation}
and the coupling coefficients
\begin{eqnarray}
&&1+\cos\theta_1\cos\theta_2 = \frac{2\delta_2^2 + \Omega_2^2 - \delta_2\Delta}{\delta_2^2+\Omega_2^2}, \label{eq_cos_times_cos} \\
&&\cos\theta_1+\cos\theta_2 = \frac{2\delta_2-\Delta}{\sqrt{\delta_2^2+\Omega_2^2}}, \label{eq_cos_plus_cos} \\
&&\sin\theta_1\sin\theta_2 = \frac{\Omega_2\sqrt{\Omega_2^2 + 2\delta_2\Delta - \Delta^2}}{\delta_2^2+\Omega_2^2}. \label{eq_sin_times_sin}
\end{eqnarray}

To eliminate $\cos\theta_1 + \cos\theta_2$, we set the driving frequency $\omega_d = (\omega_1^L + \omega_2^L)/2$. The resonance condition (\ref{eq_res_cond2}), as well as the condition $\widetilde{\omega}_{1,2}\gg \omega^{xx}$, becomes $\Omega_1 = \Omega_2 \gg \omega^{xx}$. Since $|\Delta|\approx \omega^{xx}$, the effective Hamiltonian has a rather simple form
\begin{equation}
H_\mathrm{eff}''\approx \frac{\omega^{xx}}{4}\cos\phi\left[ \sigma_x^{(1)}\sigma_x^{(2)} + \sigma_y^{(1)}\sigma_y^{(2)} + 2\sigma_z^{(1)}\sigma_z^{(2)}\right] . \label{eq_xyz_coupling}
\end{equation}

Does this switching scheme work with quantized fields as well? Here we briefly notice that in this approach it is essential to operate with states of
the electromagnetic field having well-defined phases; therefore this
switching scheme can be implemented either with classical fields or with coherent states\cite{Schleich, Scully}; Fock (number) states,
even if they could be prepared experimentally, have fluctuating phases, therefore cannot be used.

\section{Entangling properties}
\label{RWA}

In order to characterize the effectiveness of the switchable coupling schemes derived in previous section, we need to determine the on/off ratio of the coupling with the original
Hamiltonian (\ref{eq_time_dep_ham}). In this section, we will use the
{\it concurrence} \cite{Wootters} to study the entanglement between the two qubits. The concurrence of a {\it pure} two-qubit state $|\psi\rangle$ is defined as
\begin{equation}
{\cal C}(\psi) = |\langle\psi|\sigma_y\otimes\sigma_y|\psi^*\rangle| , \label{eq_conc_def1}
\end{equation}
where $|\psi^*\rangle$ is the complex conjugate of $|\psi\rangle$. For a general two-qubit state $|\psi\rangle = c_{uu}|\uparrow\uparrow\rangle + c_{ud}|\uparrow\downarrow\rangle + c_{du}|\downarrow\uparrow\rangle + c_{dd}|\downarrow\downarrow\rangle$, it is
\begin{equation}
{\cal C}(\psi) = 2|c_{uu}c_{dd}-c_{ud}c_{du}| \leq 1. \label{eq_conc_def2}
\end{equation}
We define the on/off ratio of coupling as the ratio of the maximum concurrence obtained when the coupling is
switched on to the maximum concurrence obtained when the coupling is effectively off.

In the lab frame, the concurrence can be calculated by numerically solving the Schr\"odinger equation with the
time-dependent Hamiltonian (\ref{eq_time_dep_ham}) \cite{Solving_Schrodinger_equation}. In the rotating frame, analytical calculations with the effective Hamiltonian (\ref{eq_eff_ham0}) can be done under the circumstance that $\delta_1$ and $\delta_2$ are small. Here we only focus on the zero $\delta \widetilde{\omega}$ case discussed in Sec. \ref{low_osc_freq}. For the sake of simplicity, we also consider that the driving fields satisfy $\Omega_1 = \Omega_2 \equiv \Omega \gg \omega^{xx}$, and specify our initial state to the ground state $|\uparrow\uparrow\rangle$.

We start by deriving the analytical expression of the concurrence.
When $\phi_1 = \phi_2 = 0$, the coupling is switched on. The effective Hamiltonian (\ref{eq_eff_ham0}) is approximately
\begin{equation}
H_\mathrm{eff} \approx \frac{\Omega}{2}(\sigma_1^x + \sigma_2^x) + \frac{\omega^{xx}}{2}(\sigma_1^x\sigma_2^x + \sigma_1^y
\sigma_2^y) . \label{eq_eff_ham_zero_detuning}
\end{equation}
It has eigenvalues $\lambda_1=0$, $\lambda_2=-\omega^{xx}$, $\lambda_{3,4} = [\omega^{xx} \pm \sqrt{(\omega^{xx})^2 + 4\Omega^2}]/2$, and the corresponding (unnormalized) eigenvectors $[-1,0,0,1]^\mathrm{T}$, $[0,-1,1,0]^\mathrm{T}$, $[1,\lambda_{3,4}/\Omega,\lambda_{3,4}/\Omega,1]^\mathrm{T}$, respectively. By expanding $|\uparrow\uparrow\rangle$ in terms of the eigenvectors, we get the time-dependent state vector
\begin{widetext}
\begin{eqnarray}
|\psi(t)\rangle &=& \left[ \frac{1}{4}\left( e^{-i\lambda_3t} + e^{-i\lambda_4}t + 2 \right) + \frac{\omega^{xx}}{4(\lambda_3 - \lambda_4)}\left( e^{-i\lambda_4t} - e^{-i\lambda_3t} \right) \right] |\uparrow\uparrow\rangle \nonumber \\
&& - \frac{\Omega}{2(\lambda_3 - \lambda_4)}\left( e^{-i\lambda_4t} - e^{-i\lambda_3t} \right)\left( |\uparrow\downarrow\rangle + |\downarrow\uparrow\rangle \right) \nonumber \\
&& + \left[ \frac{1}{4}\left( e^{-i\lambda_3t} + e^{-i\lambda_4}t - 2 \right) + \frac{\omega^{xx}}{4(\lambda_3 - \lambda_4)}\left( e^{-i\lambda_4t} - e^{-i\lambda_3t} \right) \right] |\downarrow\downarrow\rangle . \label{eq_time_evo_rwa1}
\end{eqnarray}
\end{widetext}
Then the concurrence is approximately
\begin{equation}
{\cal C}(t) \approx |e^{i \omega^{xx} t} -1|/2 = \sqrt{[1-\cos(\omega^{xx} t)] / 2} , \label{eq_conc_first_rwa1}
\end{equation}
oscillating between 0 and 1 with a period of $2\pi / \omega^{xx}$.

According to (\ref{eq_xyz_coupling}), if we set $\phi_1 = \pi/2$ and $\phi_2 = 0$, the coupling should be effectively switched off,
whereas the effective Hamiltonian (\ref{eq_eff_ham0}) in this case reads
\begin{equation}
H_\mathrm{eff}\approx \frac{\Omega}{2}(\sigma_2^x - \sigma_1^y) + \frac{\omega^{xx}}{2}(\sigma_1^x\sigma_2^x + \sigma_1^y\sigma_2^y) .
\end{equation}
By diagonalizing $H_\mathrm{eff}$ and expanding $|\uparrow\uparrow\rangle$ with the eigenvectors again,
we arrive at an analytical expression of the time-dependent concurrence as
\begin{equation}
{\cal C}(t) \approx \frac{(\omega^{xx})^2}{2\Omega^2}| 3 + \cos(2\Omega t) - 4\cos(\Omega t) | . \label{eq_conc_first_rwa2}
\end{equation}
The maximum concurrence is about $(2\omega^{xx})^2 / \Omega^2$.

Thus, the on/off coupling ratio is approximately $\Omega^2 / (2\omega^{xx})^2$.
To perform the numerical calculations, we set $\Omega = 10 \omega^{xx}$, $\omega_d = 200 \omega^{xx}$, $\omega_1^L = \omega_d + \Delta/2$,
and $\omega_2^L = \omega_d - \Delta/2$ (considering $\Delta \ll \omega_d$). For a charge qubit ($E_C / E_J \gg 1$), the probability of leakage
to non-computational states is negligible for such a ratio of $\Omega / \omega_{1,2}^L$ (see Appendix \ref{leakage}).

\begin{widetext}

\begin{figure}[htb]
\includegraphics[width=9cm]{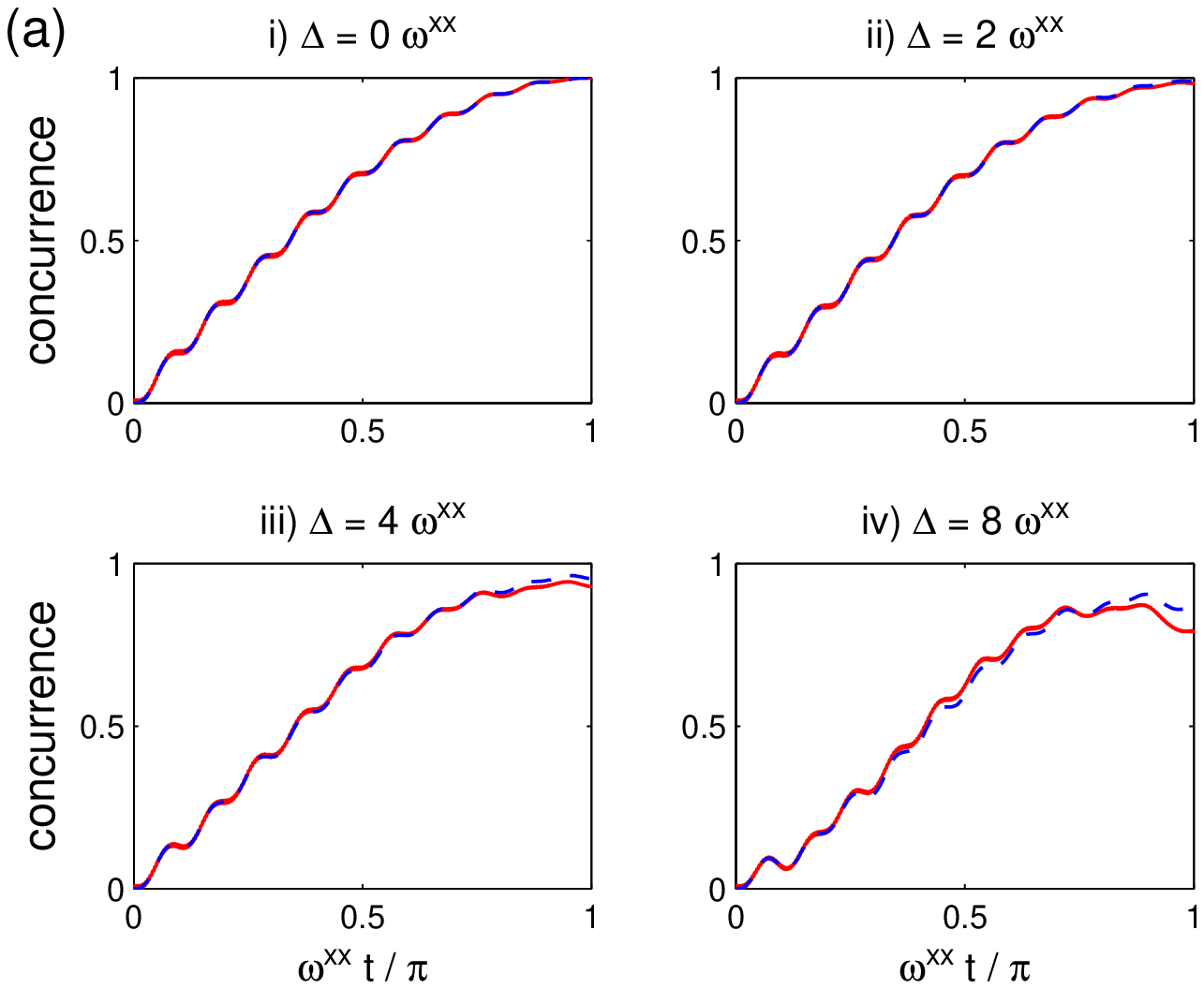}\includegraphics[width=9cm]{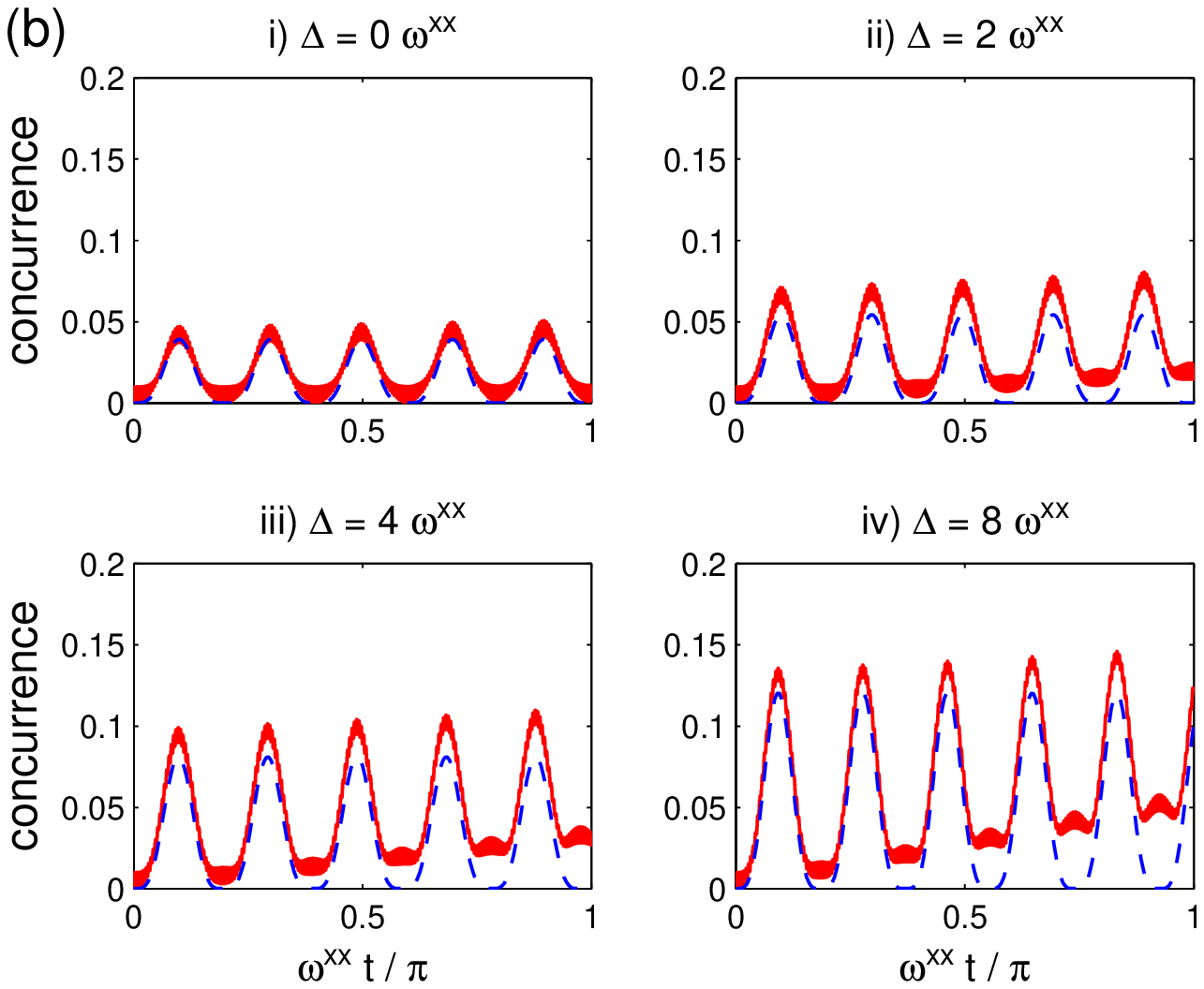}
\caption{(Color online) Time evolution of the concurrences for
the input state $|\psi_\mathrm{in}\rangle = |\uparrow\uparrow\rangle$. (a) $\phi_1 = \phi_2 = 0$; (b) $\phi_1 = \pi/2$, $\phi_2 = 0$.
The solid (red) and dashed (blue) lines indicate the concurrences calculated with the
time-dependent Hamiltonian (\ref{eq_time_dep_ham}) and the effective Hamiltonian (\ref{eq_eff_ham0}), respectively.
We have set $\Omega = 10 \omega^{xx}$, $\omega_d = 20 \Omega$, and $\omega_{1,2}^L = \omega_d \pm \Delta/2$.} \label{fig_first_rwa}
\end{figure}

\end{widetext}

In Fig. \ref{fig_first_rwa}, we show the concurrences calculated with different values of $\Delta$.
The numerical results indicate that the switchable coupling scheme works well as long as $\Delta$ is not too big compared to $\omega^{xx}$.
An on/off coupling ratio of about $20$ can be obtained in a time duration longer than several $\pi$-pulse widths.
Once $\Delta$ becomes large, the on/off ratio is reduced. Furthermore, differences between the concurrences (dashed lines)
calculated with the effective Hamiltonian (\ref{eq_eff_ham0}) and the exact concurrences (solid lines) evaluated with
the time-dependent Hamiltonian (\ref{eq_time_dep_ham}) start to appear.
These differences imply that the counter-rotating terms neglected in the RWA leading to (\ref{eq_eff_ham0}) cause effects larger than the
{Bloch-Siegert shift} \cite{Cohen,Puri}.

In order to understand how the counter-rotating terms
\begin{eqnarray}
H_\mathrm{cr} &=& e^{2i\omega_dt}\left( \omega^{xx}\sigma_1^+\sigma_2^+ + \frac{\Omega}{2} \sum_{j=1,2} e^{i\phi_j}\sigma_j^+ \right) \nonumber \\
&& + e^{-2i\omega_dt}\left( \omega^{xx}\sigma_1^-\sigma_2^- + \frac{\Omega}{2} \sum_{j=1,2} e^{-i\phi_j}\sigma_j^- \right) \ \ \ \label{eq_counter_rotating_terms}
\end{eqnarray}
behave, we extend the derivation of Bloch-Siegert shift for single two-level atom (see {\it e.g.} Chapter 7 of Ref. \cite{Puri}), to our
two-qubit system. To shorten the following discussions, we only consider $\phi_1 = \phi_2 = 0$.

The time-evolution operator generated by (\ref{eq_time_dep_ham}) can be expressed as
\begin{equation}
U(t) = S_1(t) e^{-i H_\mathrm{eff} t} \overleftarrow{T} \exp\left[ -i\int_0^t d\tau \widetilde{H}_\mathrm{cr}(\tau) \right] , \label{eq_time_evo_for_time_dep_ham}
\end{equation}
where $S_1(t)$ is defined in Eq. (\ref{eq_unit_oper1}), $H_\mathrm{eff}$ is given by (\ref{eq_eff_ham0}), $\overleftarrow{T}\exp[\cdots]$ denotes the time-ordered exponential integration, and
\begin{equation}
\widetilde{H}_\mathrm{cr}(\tau) = e^{i H_\mathrm{eff} \tau} H_\mathrm{cr} e^{-i H_\mathrm{eff} \tau} . \label{eq_counter_rotating_rwa}
\end{equation}

Because of the qubit-qubit coupling terms in $H_\mathrm{eff}$, the transformation (\ref{eq_counter_rotating_rwa}) is hard to perform. Due to the fact that only when $\delta_{1,2} = \mp\Delta / 2 \gg \omega^{xx}$ the deviations become significant, we may ignore the coupling terms, and carry out (\ref{eq_counter_rotating_rwa}) in the basis of $\{|+ +\rangle, |+ -\rangle, |- +\rangle, |- -\rangle  \}$ 
\begin{eqnarray}
\widetilde{H}_\mathrm{cr}(\tau) &\approx& \exp\left\{ \frac{i\tau}{2} \left[ \widetilde{\omega}_1 \sigma_z^{(1)} + \widetilde{\omega}_2 \sigma_z^{(2)} \right] \right\} H_\mathrm{cr}' \nonumber \\
&& \times \exp\left\{ \frac{-i\tau}{2} \left[ \widetilde{\omega}_1 \sigma_z^{(1)} + \widetilde{\omega}_2 \sigma_z^{(2)} \right] \right\} , \label{eq_tranform_counter_rotating_terms}
\end{eqnarray}
where $\widetilde{\omega}_1 = \widetilde{\omega}_2 = \sqrt{(\Delta/2)^2 + \Omega^2}$, and $H_\mathrm{cr}'$ denotes the counter-rotating terms transformed into this basis.

\begin{figure}[htb]
\includegraphics[width=9cm]{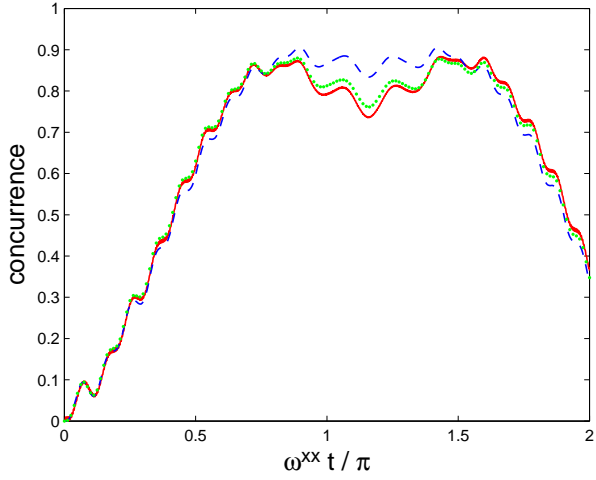}
\caption{(Color online) Time evolution of concurrences under the same values of parameters as those in Fig. \ref{fig_first_rwa}(a iv). Again, the solid (red) and dashed (blue) lines indicate the exact concurrence and the concurrence evaluated by the effective Hamiltonian with RWA. The dotted (green) line is the concurrence calculated with counter-rotating terms compensation. }
\label{fig_counter_rotating}
\end{figure}

We have evaluated the time-ordered exponential integration to second order. By transforming it back to the computational basis $\{ |\uparrow\uparrow\rangle, |\uparrow\downarrow\rangle, |\downarrow\uparrow\rangle, |\downarrow\downarrow\rangle \}$, we have obtained
\begin{eqnarray}
&& \ \ \ \overleftarrow{T}\exp\left[ -i\int_0^t d\tau \widetilde{H}_\mathrm{cr}(\tau) \right] \nonumber \\
&& \approx 1 - i\int_0^t d\tau \widetilde{H}_\mathrm{cr}(\tau) - \int_0^t d\tau_2 \int_0^{\tau_2} d\tau_1 \widetilde{H}_\mathrm{cr}(\tau_2) \widetilde{H}_\mathrm{cr}(\tau_1) \nonumber \\
&& \approx \exp\left\{ \frac{i\Omega^3 t}{\omega_d \widetilde{\omega}_1} \left[ \sigma_1^z + \sigma_2^z - \frac{\Delta}{16\widetilde{\omega}_1} \left( \sigma_1^x - \sigma_2^x \right) \right] \right\} . \label{eq_counter_rotating_shift}
\end{eqnarray}

This equation shows that besides the Bloch-Siegert shift in each qubit Larmor frequency, the counter-rotating terms also give rise
to Rabi frequency shift proportional to $\Delta$. As illustrated in Fig. \ref{fig_counter_rotating}, the concurrence recalculated
by substituting (\ref{eq_counter_rotating_shift}) in (\ref{eq_time_evo_for_time_dep_ham}) (dotted line) fits the exact concurrence (solid line) much better than the concurrence calculated barely with $H_\mathrm{eff}$ (dashed line).

\section{Implementation of quantum gates by composite pulses}
\label{quantum_gates}

We are now in the position to address the problem of implementing quantum gates. Simple rectangular driving pulses have been proved to be
not satisfactory for precise manipulations \cite{Steffen}. It is better to adopt other pulse techniques \cite{Vandersypen} such as shaped
pulses and composite pulses. In order to achieve arbitrary quantum gates with both high accuracy and high speed, we need to
find a numerical approach to get the suitable pulse parameters (amplitudes, frequencies, phases and pulse durations). The idea is to find a certain functional, and to obtain the optimal pulse parameters by maximizing or minimizing this functional. In a previous work \cite{Paraoanu}, we found a set of parameters to get a locally equivalent gate
of CNOT by minimizing the functional $|G_1(t)|^2+|G_2(t)-1|^2$, where $G_1$ and
$G_2$ are two locally invariant functionals defined by Makhlin \cite{Makhlin}. This method has a drawback that, when a locally equivalent gate
is found, determining the two single-qubit gates which transform it into CNOT could be a difficult task. Therefore in this section, we minimize
a different functional \cite{Niskanen} which can lead us directly to the target gate we want.

We work in the rotating frame to reduce the computational complexity. Since in the preceding section we have found that the RWA broke down when $\Delta$ was large, here we assume $\Delta = 2\omega^{xx}$, and set the detunings $\delta_1$ and $\delta_2$ to be $-\Delta/2$ and $\Delta/2$, respectively. The pulse duration of performing a gate, $t_p$, is equally divided into $m$ small intervals. We consider $\phi_1$ and $\phi_2$ as \lq\lq global\rq\rq parameters which are unchanged in the whole pulse duration. In order to avoid sharp edges as in rectangular pulses, we assume that in each time interval $dt$ the \lq\lq local\rq\rq pulse parameters $\Omega_{1,2}$ vary linearly with time, and at $t = 0$ and $t = t_p$, $\Omega_{1,2} = 0$. The unitary operator generated by the effective Hamiltonian (\ref{eq_eff_ham0}) right after the pulses can be well approximated as
\begin{widetext}\begin{eqnarray}
U_\mathrm{eff} \approx
\exp\left\{-idtH_{\mathrm{eff}}\left[\frac{\Omega_1^{(m)} + \Omega_1^{(m-1)}}{2}, \frac{\Omega_2^{(m)} + \Omega_2^{(m-1)}}{2}\right] \right\} \cdots
\exp\left\{-idtH_{\mathrm{eff}}\left[\frac{\Omega_1^{(1)} + \Omega_1^{(0)}}{2}, \frac{\Omega_2^{(1)} + \Omega_2^{(0)}}{2}\right] \right\}, \label{eq_app_univ_oper}
\end{eqnarray}\end{widetext}
where $\Omega_{1,2}^{(k)}$ ($k > 0$) stand for the values of $\Omega_{1,2}$ at the end of the $k$th interval, and $\Omega_{1,2}^{(0)}$ denote $\Omega_{1,2}$ at $t = 0$. The optimal $\phi_{1,2}$, $\Omega_{1,2}^{(k)}$ and $t_p$ for a target gate $U$ are achieved by searching for the global minimum of the error functional
\begin{equation}
\epsilon = \sqrt{\mathrm{Tr}[(U - U_\mathrm{eff})^\dag(U - U_\mathrm{eff})]}. \label{eq_frob_norm}
\end{equation}

\subsection{Simulated annealing}
\label{simulated_anneal}

Since normally $\epsilon$ has many local minima, to avoid being trapped in a local minimum, we use the {\it simulated
annealing} (SA) method \cite{Kirkpatrick, Press} to minimize $\epsilon$. It employs a random search of pulse parameters which allows
changes not only decreasing the \lq\lq energy\rq\rq $\epsilon$, but also increasing it. The probability to accept a change is given by $P = \exp(-\delta\epsilon / T)$, where $\delta\epsilon = \epsilon_\mathrm{after\ the\ change} - \epsilon_\mathrm{before\ the\ change}$, and $T$ is a parameter acting as the system \lq\lq temperature\rq\rq.
One can easily find that, if a change of pulse parameters
decreases $\epsilon$, $P$ is always larger than unity, which means we
always accept this change; if a change increases $\epsilon$, we still have
chance to accept this change. Our algorithm is summarized as
follows.

(i) Define the starting temperature $T_\mathrm{s} = -0.01 / \ln(0.8)$ and
the halting temperature $T_\mathrm{h} = -10^{-10} / \ln(0.8)$ for the annealing process.

(ii) Define the boundaries of pulse parameters $\Omega_{1,2}^{(k)}\in[0,10\omega^{xx}]$ and $\phi_{1,2}\in[-\pi/2,\pi/2]$.

(iii) Initialize the pulse parameters $\Omega_{1,2}^{(k)} = 5\omega^{xx}$ and $\phi_{1,2} = 0$. Calculate the initial value of $\epsilon$ with these pulse parameters. Initialize the temperature $T = T_\mathrm{s}$.

(iv) Repeat the following steps if $T > T_\mathrm{h}$:

{\it a.} randomly generate sequences of pulse parameters until all
the parameters in the sequence are inside the boundaries;

{\it b.} evaluate $\epsilon$ with the last sequence;

{\it c.} keep the resulting $\epsilon$ and the pulse sequence with a
probability $P=\exp(-\delta\epsilon / T)$;

{\it d.} after every 1000 successful evaluations of $\epsilon$, lower the
temperature by $1\%$.

We demonstrate two examples here: a $\pi$-rotation of the first qubit around its $X$-axis
\begin{equation}
U_{X_1} = e^{-i\sigma_1^x\pi/2}\otimes I_2 = \left[
\begin{array}{cccc}
        0 & 0 & -i & 0 \\
        0 & 0 & 0 & -i \\
        -i & 0 & 0 & 0 \\
        0 & -i & 0 & 0
        \end{array}
    \right] , \label{eq_x1}
\end{equation}
and a CNOT gate
\begin{equation}
U_{\mathrm{CNOT}} = e^{-i\pi/4}\left[
    \begin{array}{cccc}
        1 & 0 & 0 & 0 \\
        0 & 1 & 0 & 0 \\
        0 & 0 & 0 & 1 \\
        0 & 0 & 1 & 0
        \end{array}
    \right],
    \label{eq_cnot}
\end{equation}
where the factor $\exp(-i\pi/4)$ is used to make $U_{\mathrm{CNOT}}\in \mathrm{SU}(4)$, since $\mathrm{Tr}(H_{\mathrm{eff}}) = 0$.

\begin{figure}[htb]
\includegraphics[width=9cm]{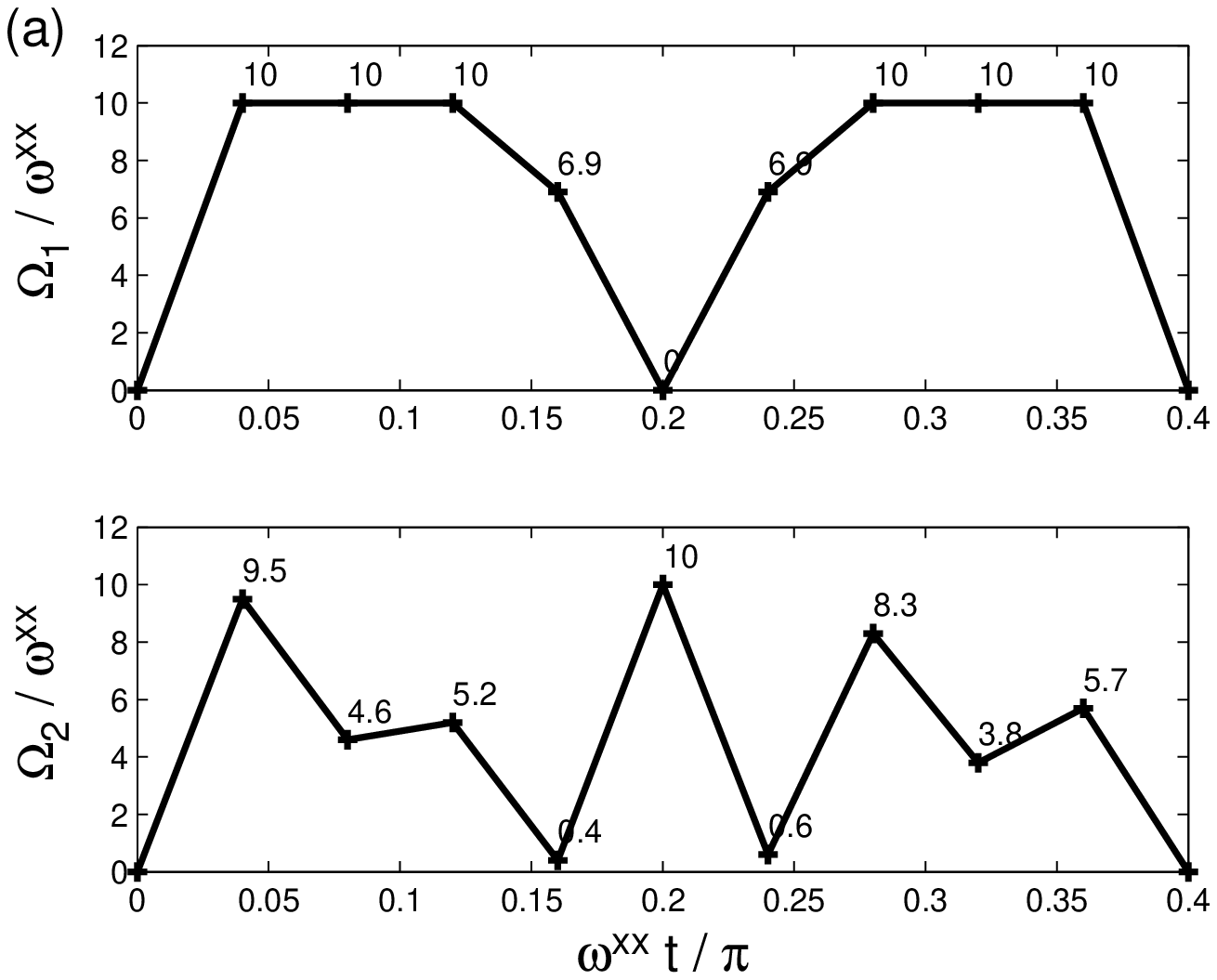}
\includegraphics[width=9cm]{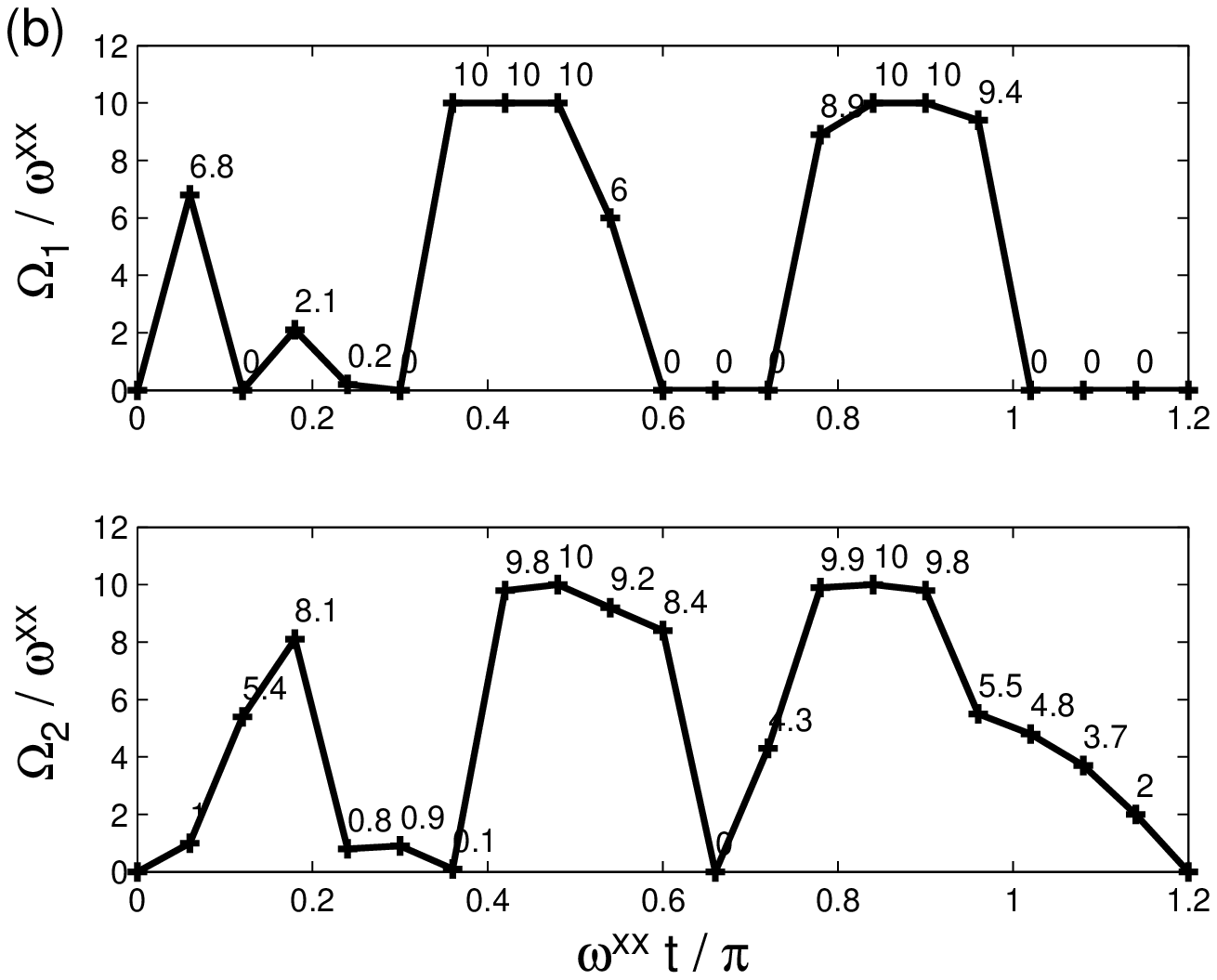}
\caption{Possible sequences of $\Omega_1(t)$ and $\Omega_2(t)$ (a) for the single-qubit gate $X_1$, (b) for the CNOT gate. }
\label{fig_gate}
\end{figure}

Because of the off-resonance shaped pulses and the finite on/off coupling ratio, a pulse duration much longer than $\pi/\max(\Omega) = 0.1\pi / \omega^{xx}$ was expected for $X_1$ gate. By running our SA program numerous times, we have found it very difficult to minimize $\epsilon$ to a satisfactory value if $t_p < 0.4\pi / \omega^{xx}$. In Fig. \ref{fig_gate}(a), a possible control sequence of $\{\Omega_1(t),\Omega_2(t)\}$ is presented. The pulse duration $t_p = 0.4\pi / \omega^{xx}$ has been divided into 10 intervals, and the optimized phases $\phi_1 = 0$, $\phi_2 = -0.5\pi$.

The minimum pulse duration for the CNOT gate we have found so far is $t_p = 1.2\pi / \omega^{xx}$. Interestingly,
it is approximately equal to the summation of the {\it interaction cost} \cite{Vidal} $C_H(\mathrm{CNOT}) = 0.5\pi / \omega^{xx}$ for
our effective Hamiltonian, and durations for two single-qubit gates, which indicates that the SA program
constructs the CNOT gate in a way close to a Cartan decomposition \cite{Cartan}. A possible sequence of $\{\Omega_1(t),\Omega_2(t)\}$ is shown in Fig. \ref{fig_gate}(b). The corresponding phases $\phi_1 = -0.2\pi$ and $\phi_2 = 0.07\pi$.

\subsection{Obtaining maximally entangled two-qubit states}

\begin{figure}[htb]
\includegraphics[width=9cm]{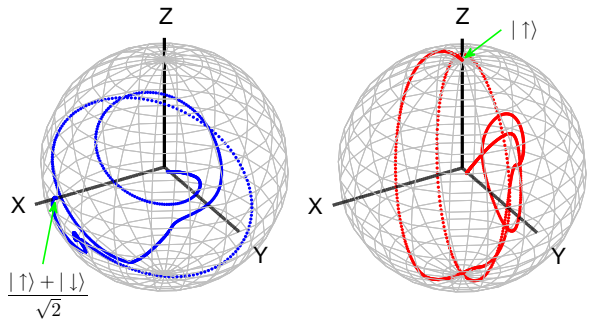}
\caption{(Color online) Trajectories of the reduced density matrices in the rotating reference frame, projected onto the Bloch spheres.
The blue and the red dots denote the motions of qubit-1 and qubit-2, respectively. }
\label{fig_bloch-sphere}
\end{figure}

To get a visual picture of how these optimized driving pulses work, we perform a simulation of producing a maximally
entangled two-qubit state by the CNOT gate. Suppose that initially the two qubits were in the ground state $|\uparrow\uparrow\rangle$ and
then
a $\pi/2$-rotation around $Y$-axis of qubit-1 has been applied;
so the new input state is then
\begin{equation}
|\psi_\mathrm{in}\rangle = \frac{|\uparrow\rangle + |\downarrow\rangle}{\sqrt{2}}\otimes |\uparrow\rangle ,
\end{equation}
as indicated with arrows in Fig. \ref{fig_bloch-sphere}.

We then send the qubits through a CNOT gate, realized using the pulse sequences presented in Fig. \ref{fig_gate}(b); then
the output state of the two qubits will be a {\it Bell state} \cite{Nielsen}
\begin{equation}
|\beta_{\uparrow\uparrow}\rangle = \frac{1}{\sqrt{2}} ( |\uparrow\uparrow\rangle + |\downarrow\downarrow\rangle ) . \label{eq_bell_state}
\end{equation}

The total density matrix for this system of two coupled qubits is obtained by numerically solving the Schr\"odinger equation with
the effective Hamiltonian (\ref{eq_eff_ham0}); the reduced density matrix of each single qubit is obtained by partially tracing
out qubit-1 or qubit-2. Fig. \ref{fig_bloch-sphere} shows the motions of the reduced density matrices in the Bloch sphere picture.
The reduced density matrices of each qubit
end up in the centers of each of the spheres (corresponding to a zero Bloch vector),
indicating that the two qubits are in a maximally entangled state \cite{Nielsen}.

For the total density matrix of the output state in rotating frame $\rho_\mathrm{out}$, we find
a state fidelity \cite{Vandersypen}
\begin{equation}
{\cal F}(|\beta_{\uparrow\uparrow}\rangle,\rho_\mathrm{out}) = \sqrt{\langle \beta_{\uparrow\uparrow}|\rho_\mathrm{out}|\beta_{\uparrow\uparrow} \rangle} > 0.999 .
\end{equation}

\begin{figure}[hbt]
\includegraphics[width=9cm]{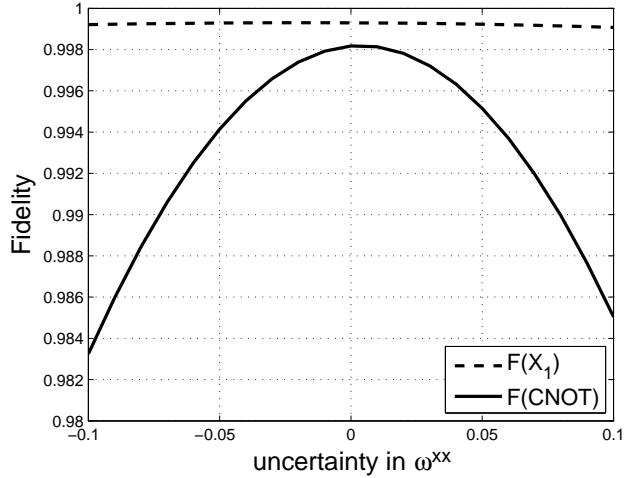}
\caption{The fidelity loss due to uncertainty in measuring $\omega^{xx}$. }
\label{fig_fidelity_vs_em}
\end{figure}

\subsection{Losses of gate fidelity}

Now, we move back to the lab frame to evaluate the gate fidelity \cite{Fidelity}
\begin{equation}
{\cal F}(U) \equiv \overline{\langle\psi_\mathrm{in}| U^\dag \rho_\mathrm{out}U|\psi_\mathrm{in}\rangle} \label{eq_define_fidelity}
\end{equation}
with the time-dependent Hamiltonian (\ref{eq_time_dep_ham}). We set $\omega_d = 200\omega^{xx}$
and $\omega_{1,2}^L = \omega_d - \delta_{1,2}$, as in Sec. \ref{RWA}. By numerically solving the Schr\"odinger equation,
we obtain ${\cal F}(X_1) = 0.9993$ and ${\cal F}(\mathrm{CNOT}) = 0.9982$. Such high fidelities can only be
obtained in rather ideal cases, since the calculations have only taken gate errors due to the counter-rotating terms into account,
and these errors are very small by choosing small qubit-qubit detuning. In practice, uncertainties in system parameters, external noises,
etc. will cause extra fidelity losses.

\begin{figure}[htb]
\includegraphics[width=9cm]{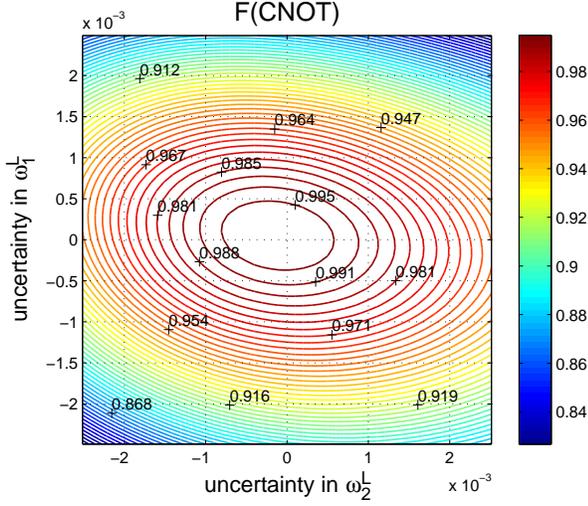}
\caption{(Color online) Loss of ${\cal F}(\mathrm{CNOT})$ due to the uncertainties in measuring $\omega_{1,2}^L$. ${\cal F}(X_1)$ has similar features to ${\cal F}(\mathrm{CNOT})$, but less sensitive to the uncertainties (data not shown). }
\label{fig_fidelity_loss_ej}
\end{figure}

The optimization method we have used relies on the well defined qubit parameters,
such as the Larmor frequencies $\omega_{1,2}^L$ and coupling strength $\omega^{xx}$. In experiments on Josephson qubits,
although the qubit parameters may be tunable due to specific designs, we still assume they are static, and the detailed knowledge
of them are obtained by measurements. So, uncertainties in measurements of qubit Larmor frequencies and coupling strength will give rise to
gate errors. By performing numerical simulations, we have found that, compared with ${\cal F}(\mathrm{CNOT})$, ${\cal F}(X_1)$ is less sensitive
to the uncertainties in qubit parameters. As shown in Fig. \ref{fig_fidelity_vs_em}, there is nearly no loss of ${\cal F}(X_1)$ even
when the measured value of $\omega^{xx}$ is $90\%$ (or $110\%$) of the exact $\omega^{xx}$, whereas in order to keep
high ${\cal F}(\mathrm{CNOT})$ (say $> 0.99$), the uncertainty in $\omega^{xx}$ should not be more than $7\%$.
The data in Fig. \ref{fig_fidelity_loss_ej} indicate that the uncertainties in measuring $\omega_{1,2}^L$ should be controlled within $0.05\%$,
which is easily achievable with present-day electronics.
If we consider typical charge qubits and dispersive coupling through a resonator, $\omega_{1,2}^L \approx 2\pi \times 5$ GHz and
$\omega^{xx} \approx 2\pi \times 20$ MHz (see \cite{yale2}), the allowed uncertainties are in MHz range.

Nevertheless, the uncertainties in the driving pulses are not troublesome to the optimized gates. As shown in Sec. \ref{simulated_anneal},
the required precision of pulse parameters are already low, only $0.1\omega^{xx}$ for $\Omega_{1,2}$ and $0.01\pi$ for $\phi_{1,2}$. To account for experimental imperfections
(jitter and phase noise of the external electronics used to create the pulses), we also simulate the errors in pulses as normally
distributed random numbers with relative error $\sigma$ and null average, and find that if $\sigma$ is smaller than $1\%$, the loss of fidelity is negligible.

\begin{figure}[htb]
\includegraphics[width=9cm]{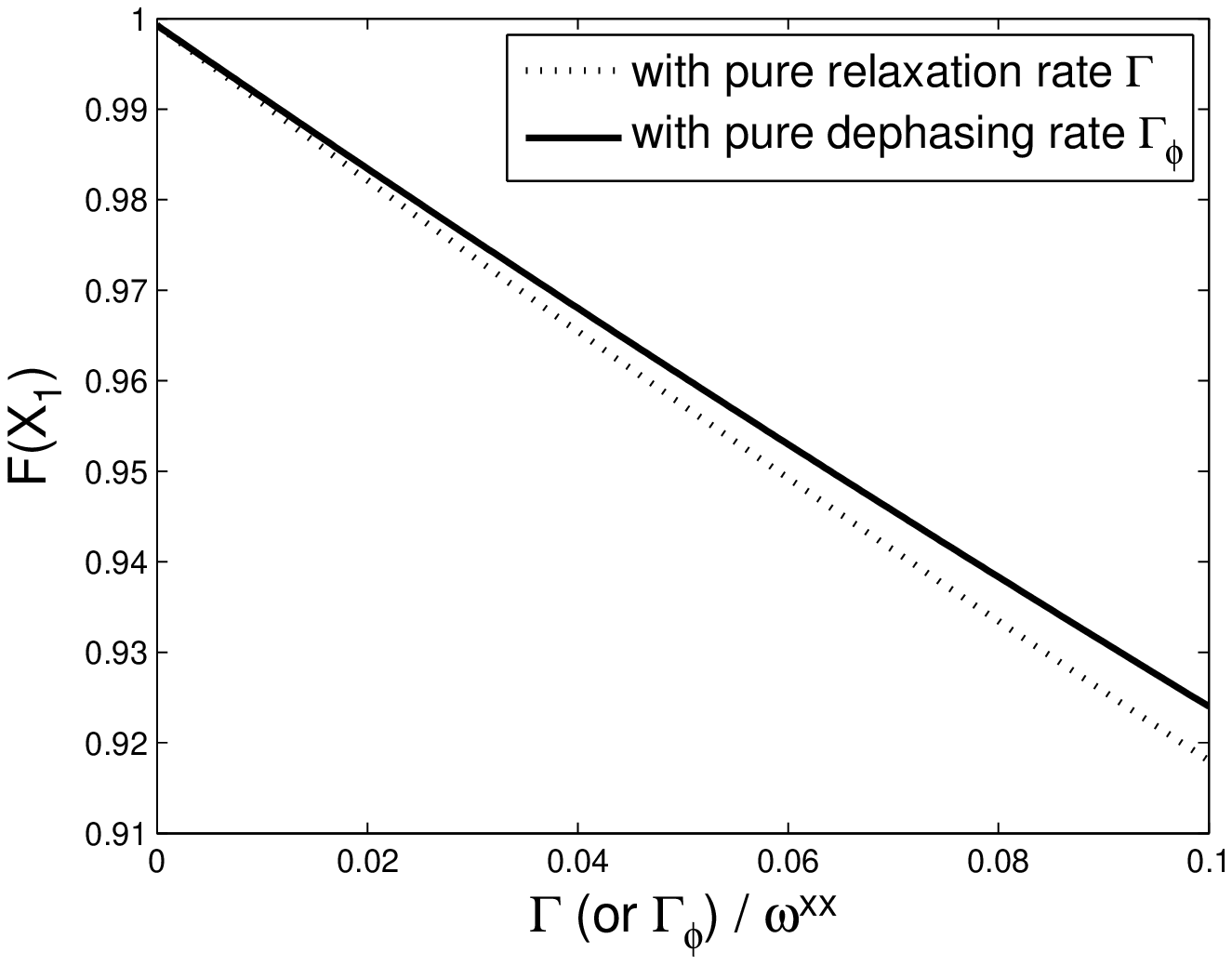}
\includegraphics[width=9cm]{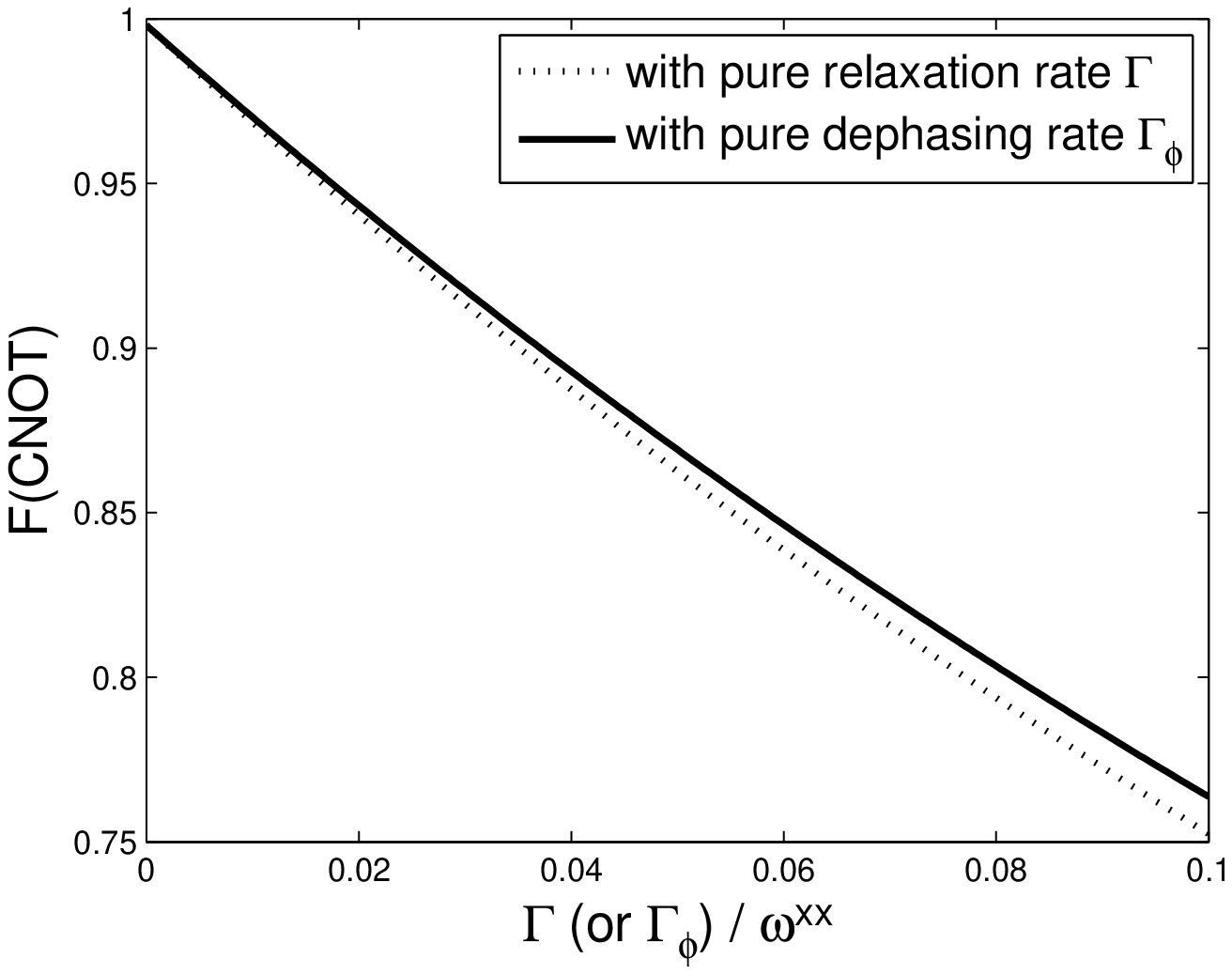}
\caption{Loss of two-qubit (CNOT) fidelity ${\cal F}(\mathrm{CNOT})$
and single-qubit ($X_{1}$) fidelity ${\cal F}(X_1)$ due to decoherence.}
\label{decoherence}
\end{figure}

We now examine the effect of the electromagnetic degrees of freedom which inevitably couple to each of the two qubits, producing decoherence.
We consider here the worst-case scenario \cite{Li}, in which each qubit is coupled to a different environment,
modeled by Lindblad superoperators
\begin{eqnarray}
{\cal L}_{tj}[\rho] = \frac{\Gamma_\phi^{(j)}}{2}\left( \sigma_j^z\rho\sigma_j^z - \rho \right) ,
\end{eqnarray}
and
\begin{eqnarray}
{\cal L}_{lj}[\rho] &=& \frac{\Gamma^{(j)}}{2} \left( 2\sigma_j^-\rho\sigma_j^+ - \sigma_j^+\sigma_j^-\rho - \rho\sigma_j^+\sigma_j^- \right),
\end{eqnarray}
describing longitudinal and transversal noise with decay rates $\Gamma^{(j)}$ and $\Gamma^{(j)}_{\phi }$ respectively.
We evolve the system numerically under the
simultaneous action of decoherence
 and of the pulse sequences corresponding to single
gates and two-qubit gates.  To simplify the presentation we take the decoherence  rates of the two qubits equal,
$\Gamma^{(j)}=\Gamma$, $\Gamma^{(j)}_{\phi }=\Gamma_{\phi }$, and we
show the resulting fidelity loss due to decoherence in Fig. \ref{decoherence}.

\subsection{Summary}

Compared to the FLICFORQ protocol developed in \cite{Rigetti}, our protocol do improve the speed of two-qubit gate, due to the fact
that in our schemes the qubit-qubit coupling strength is reduced only by a factor of 2, and it still holds the
advantages of FLICFORQ: no need to dc bias away from the optimal points and no need for active tunable coupling. For a relatively
weak coupling mediated through cavity, $\omega^{xx} \approx 2\pi \times 20$ MHz, a CNOT gate can still be performed in about 30 ns.
With two qubits directly coupled by a capacitor, $\omega^{xx}$ can be much larger, and it is only constrained by $\omega^{xx}\ll \Omega_{1,2}\ll \omega_{1,2}^L$.

We have only demonstrated quantum gates with the switchable scheme for low $\delta\widetilde{\omega}$ (see Sec. \ref{low_osc_freq}),
however, the numerical optimization method used in this section should be applicable also to the high $\delta\widetilde{\omega}$
case (in Sec. \ref{high_osc_freq}), with one more pulse parameter $\omega_d$ needed to be optimized. Since the high $\delta\widetilde{\omega}$
scheme can be realized by single-qubit driving, there might be possible applications other than manipulating qubits, like bringing a qubit on and off resonance with a naturally-formed two level system (TLS) located in the Josephson junction barrier, as briefly discussed in Appendix \ref{qubit_tls}.

\section{Quantum effects: QND entanglement}
\label{QND}

In the preceding sections, we considered that the coupling strengths between the qubits and the resonator
 are much smaller than the detunings between them. Two-qubit entanglement can be realized by using the CNOT gate. 
 In this section, we move
 to the non-dispersive regime, and propose a protocol
  to create maximally entangled two-qubit states using a method \cite{Plenio} inspired from cavity QED systems.

We still bias the dc gate voltages of the qubits at charge degeneracy points. After exciting one qubit, we switch off the driving fields.
The Hamiltonian of this qubit-qubit-resonator system is then
\begin{eqnarray}
H = -\sum_{j=1,2}\frac{E_{Jj}}{2}\sigma_j^z + \omega_r a^\dag a + i\sum_{j=1,2}g_j(a^\dag \sigma_j^- - a \sigma_j^+) . \nonumber
\end{eqnarray}
By considering that the two qubits and the resonator have the same energy  $E_{J1} = E_{J2} = \omega_r$, the Hamiltonian projected
onto the basis states $|\tilde{1}\rangle \equiv |\downarrow\uparrow\rangle \otimes |0\rangle_\mathrm{p}$,
$|\tilde{2}\rangle \equiv |\uparrow\downarrow\rangle \otimes |0\rangle_\mathrm{p}$ and
$|\tilde{3}\rangle \equiv |\uparrow\uparrow\rangle \otimes |1\rangle_\mathrm{p}$, with
$|0\rangle_\mathrm{p}$ and $|1\rangle_\mathrm{p}$ the zero-photon and one-photon Fock states of the field in resonator, has the form
\begin{equation}
H = \left[
    \begin{array}{ccc}
        0 & 0 & -i g_1 \\
        0 & 0 & -i g_2 \\
        i g_1 & i g_2 & 0
        \end{array}
    \right] .
\end{equation}

It has eigenvalues
\begin{equation}
\lambda_{1,2} = \pm \sqrt{g_1^2 + g_2^2} , \ \ \ \ \lambda_3 = 0, \label{eq_eigenvalue}
\end{equation}
with corresponding eigenvectors
\begin{eqnarray}
|\lambda_{1,2}\rangle &=& \mp \frac{i}{\sqrt{2 (g_1^2 + g_2^2)}} \left( g_1 |\tilde{1}\rangle + g_2 |\tilde{2}\rangle \right) + \frac{1}{\sqrt{2}} |\tilde{3}\rangle , \nonumber \\
|\lambda_3\rangle &=& \frac{1}{\sqrt{g_1^2 + g_2^2}} \left( -g_2 |\tilde{1}\rangle + g_1 |\tilde{2}\rangle \right) . \label{eq_eigenvector}
\end{eqnarray}
For symmetric couplings $g_1 = g_2 \equiv g$, the eigenstate $|\lambda_3\rangle$ is a direct product of the resonator vacuum state and a maximally entangled two-qubit state.

An arbitrary initial state $|\psi_\mathrm{in}\rangle = a_1|\tilde{1}\rangle + a_2|\tilde{2}\rangle + a_3|\tilde{3}\rangle$ ($|a_1|^2 + |a_2|^2 + |a_3|^2 = 1$) can be rewritten as
\begin{eqnarray}
|\psi_\mathrm{in}\rangle &=& \left[ \frac{i(a_1 + a_2)}{2} + \frac{a_3}{\sqrt{2}} \right] |\lambda_1\rangle - \frac{a_1 - a_2}{\sqrt{2}} |\lambda_3\rangle \nonumber \\
&& + \left[ -\frac{i(a_1 + a_2)}{2} + \frac{a_3}{\sqrt{2}} \right] |\lambda_2\rangle , \label{eq_ini_state2}
\end{eqnarray}
and the time evolution of it
\begin{widetext}
\begin{eqnarray}
|\psi(t)\rangle &=& e^{-i\lambda_1t}\left[ \frac{i(a_1 + a_2)}{2} + \frac{a_3}{\sqrt{2}} \right] |\lambda_1\rangle + e^{-i\lambda_2 t}\left[ -\frac{i(a_1 + a_2)}{2} + \frac{a_3}{\sqrt{2}} \right] |\lambda_2\rangle - e^{-i\lambda_3 t}\frac{a_1 - a_2}{\sqrt{2}} |\lambda_3\rangle \nonumber \\
&=& \left[ \frac{a_1-a_2}{2} + \frac{1}{2}\left( \frac{a_1+a_2}{2} - \frac{ia_3}{\sqrt{2}} \right) e^{-i\sqrt{2}gt} + \frac{1}{2}\left( \frac{a_1+a_2}{2} + \frac{ia_3}{\sqrt{2}}\right)e^{i\sqrt{2}gt} \right] |\tilde{1}\rangle \nonumber \\
&& + \left[ \frac{a_2-a_1}{2} + \frac{1}{2}\left( \frac{a_1+a_2}{2} - \frac{ia_3}{\sqrt{2}} \right)e^{-i\sqrt{2}gt} + \frac{1}{2}\left( \frac{a_1+a_2}{2} + \frac{ia_3}{\sqrt{2}}\right)e^{i\sqrt{2}gt} \right] |\tilde{2}\rangle \nonumber \\
&& + \left\{ \left[ \frac{i(a_1+a_2)}{2\sqrt{2}} + \frac{a_3}{2} \right] e^{-i\sqrt{2}gt} + \left[ -\frac{i(a_1+a_2)}{2\sqrt{2}} + \frac{a_3}{2}\right] e^{i\sqrt{2}gt} \right\} |\tilde{3}\rangle \nonumber \\
&\equiv& c_1(t)|\tilde{1}\rangle + c_2(t)|\tilde{2}\rangle + c_3(t)|\tilde{3}\rangle . \label{eq_state_time_evo1}
\end{eqnarray}
\end{widetext}

\begin{figure}[htb]
\includegraphics[width=9cm]{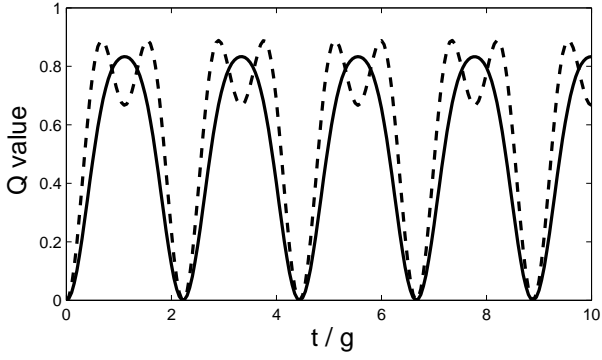}
\caption{The $Q$ value as a function of time. The solid line is for an initial excitation in one of the qubits, and the dashed line for an initial excitation in the cavity. }
\label{fig_q_measure}
\end{figure}

The global entanglement among the two qubits and one photon can be quantified by the {\it Q-measure} \cite{Meyer}
\begin{equation}
Q(t) = \frac{8}{3}\left[|c_1(t)c_2(t)|^2 + |c_1(t)c_3(t)|^2 + |c_2(t)c_3(t)|^2 \right] , \label{eq_q_measure}
\end{equation}
which can be analytically calculated with Eq. (\ref{eq_state_time_evo1}).
In Fig. \ref{fig_q_measure}, numerical calculations of the $Q$ value are illustrated.

Now, we consider the consequences of the fact that the resonator is lossy: we look at the case in which, by strongly outcoupling the
cavity to a detector (e.g. the CBJJ of Fig. \ref{fig_system})
the decay rate of the cavity is engineered to be much larger
than the energy relaxation rate of each qubit.

By continuously monitoring the cavity in a period longer than the cavity life time but much shorter than the qubit relaxation time,
we can know the state of the two qubits. If a photon is emitted out the cavity,
the system collapses to the ground state $|\tilde{0}\rangle \equiv |\uparrow\uparrow\rangle \otimes |0\rangle_\mathrm{p}$.
Then we start from the beginning by re-exciting one qubit and repeating the monitoring.
If no photon is detected, the system is in the eigenstate $|\lambda_3\rangle$, which means a two-qubit entangled state is prepared.

Experimentally,  the monitoring can be done by biasing the CBJJ to be in resonance with the resonator, $|\omega_{10} - \omega_r| \ll \kappa$, where
\begin{equation}
\kappa \approx \omega_{10}\sqrt{\frac{C_{m3}^2}{2 C_{\Sigma 3}L\tilde{c}}}
\end{equation}
is the coupling strength between the CBJJ and the resonator, and the coupling has the form (see Eq. (\ref{eq_ham_r-c}))
\begin{equation}
H_\mathrm{R-C} \approx - \kappa (a^\dag\sigma_3^- + a\sigma_3^+) .
\end{equation}

With an asymmetric design of the stripline resonator \cite{Houck}, such
that the line has only one coupling capacitor $C_{m3}$ at one end, the total decay rate of the cavity is approximately equal to $\kappa$.

\begin{figure}[htb]
\includegraphics[width=8cm]{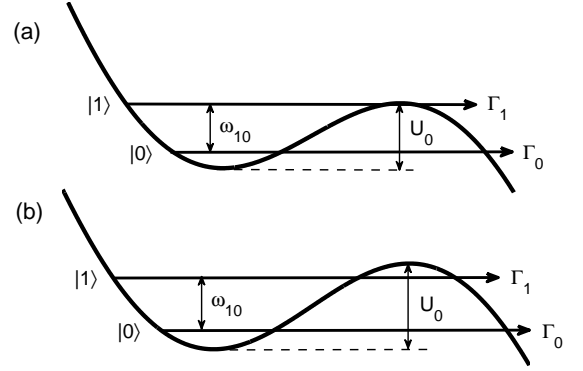}
\caption{Schematic potential energy diagram of the CBJJ (phase regime). }
\label{fig_washboard}
\end{figure}

In order to make sure that the photon absorbed by the CBJJ will never go back to the cavity,
the macroscopic quantum tunneling (MQT) rate $\Gamma_1$ for the excited state $|1\rangle$ of the CBJJ should be much larger than $\kappa$ \cite{note2}.
Thus the barrier height $U_0$ of the CBJJ should be close to $1.5 \omega_{10}$, as shown in Fig. \ref{fig_washboard}(a).
When a CBJJ is excited by absorbing a photon, it
immediately switches to the dissipative branch and creates a voltage pulse which can be measured
easily either directly or by the use of an additional dc Squid \cite{Devoret}.

The MQT rate for the ground state $|0\rangle$ in this case can be calculated as \cite{Likharev}
\begin{equation}
\Gamma_0 = \frac{\omega_p}{2\pi}\sqrt{\frac{864\pi U_0}{\omega_p}} \exp\left( -\frac{36 U_0}{5\omega_p} \right)
\approx 2 \times 10^{-4} \omega_{10} ,
\end{equation}
by assuming the plasma frequency $\omega_p \approx \omega_{10}$. The rate $\Gamma_1$ is about 500 times of $\Gamma_0$.

To achieve the desired effect with good enough efficiency, the cavity decay rate $\kappa$ should be in the range
of $\Gamma_0 \ll \kappa \ll \Gamma_1$.

This method is also suitable for the short resonator case. As derived in Appendix \ref{capacitor_coupling}, the Hamiltonian for the qubits-CBJJ system reads
\begin{eqnarray}
H = -\sum_{j=1,2}\frac{E_{Jj}}{2}\sigma_j^z - \frac{\omega_{10}}{2}\sigma_3^z + \frac{E_{12}}{4}\sigma_1^x\sigma_2^x - \sum_{j=1,2} \frac{\gamma_j}{2}\sigma_j^x\sigma_3^y . \nonumber
\end{eqnarray}
If the coupling capacitance $C_{m3}\gg C_{m1,2}$, the direct coupling $E_{12}$ between the two qubits is negligible,
and the CBJJ itself acts as a lossy cavity. By biasing the CBJJ on resonance with the qubits, the Hamiltonian projected onto the states $|\tilde{1}\rangle = |\downarrow\uparrow\rangle \otimes |0\rangle$, $|\tilde{2}\rangle = |\uparrow\downarrow\rangle \otimes |0\rangle$ and $|\tilde{3}\rangle = |\uparrow\uparrow\rangle \otimes |1\rangle$ is then
\begin{equation}
H = \frac{1}{2} \left[
    \begin{array}{ccc}
        0 & 0 & i \gamma_1 \\
        0 & 0 & i \gamma_2 \\
        -i \gamma_1 & -i \gamma_2 & 0
        \end{array}
    \right] ,
\end{equation}
which has eigenvalues and eigenvectors similar to those in Eqs. (\ref{eq_eigenvalue}) and (\ref{eq_eigenvector}).

In this case, the cavity decay rate is given by the MQT rate of the upper level $\Gamma_1$. It has to be much larger than the
qubit decay rates, but not necessarily to be as large as that in the long resonator case. So we can bias the CBJJ barrier $U_0$ to be
higher than $1.5\omega_{10}$. As illustrated in Fig. \ref{fig_washboard}(b), when $U_0 \approx 1.84 \omega_{10}$, $\Gamma_1$ is approximately
equal to $0.01\omega_{10}$, which can be of the same order as the couplings $\gamma_j$.

To study the conditional time evolution of the system state, we introduce a non-Hermitian Hamiltonian \cite{Carmichael}
\begin{equation}
H_\mathrm{cond} = H - i\Gamma \sum_{j=1,2} \sigma_j^+ \sigma_j^- - i\Gamma_1 \sigma_3^+\sigma_3^- . \label{eq_cond_ham}
\end{equation}
Here we assume that both qubits are equally coupled to the CBJJ $\gamma_1 = \gamma_2 \equiv \gamma$, and have the same relaxation rate $\Gamma \ll \Gamma_1,\ \gamma$.

The eigenvalues and corresponding eigenvectors of $H_\mathrm{cond}$ are
\begin{equation}
\lambda_{1,2}' \approx -\frac{1}{2}\left( i\Gamma_1 \pm \sqrt{2\gamma^2 - \Gamma_1^2} \right) , \ \ \lambda_3' = -i \Gamma ,
\end{equation}
and
\begin{eqnarray}
|\lambda_{1,2}'\rangle &\approx & - \frac{\Gamma_1 \pm i\sqrt{2\gamma^2 - \Gamma_1^2}}{2\sqrt{2}\gamma} (|\tilde{1}\rangle + |\tilde{2}\rangle) + \frac{1}{\sqrt{2}}|\tilde{3}\rangle , \nonumber \\
|\lambda_3'\rangle &=& -\frac{1}{\sqrt{2}}( |\tilde{1}\rangle - |\tilde{2}\rangle ) .
\end{eqnarray}

For an initial state, say,
\begin{equation}
|\psi(0)\rangle = |\tilde{1}\rangle \approx \frac{i\gamma (|\lambda_1'\rangle - |\lambda_2'\rangle) }{\sqrt{4\gamma^2 - 2\Gamma_1^2}} - \frac{|\lambda_3'\rangle}{\sqrt{2}} ,
\end{equation}
the (unnormalized) state vector at later time $t$ is given by
\begin{eqnarray}
|\psi(t)\rangle &\approx & \frac{i\gamma}{\sqrt{4\gamma^2 - 2\Gamma_1^2}} \left( e^{-i\lambda_1' t} |\lambda_1'\rangle - e^{-i\lambda_2' t} |\lambda_2'\rangle \right) \nonumber \\
&& - \frac{1}{\sqrt{2}}e^{-i\lambda_3' t} |\lambda_3'\rangle . \label{apr}
\end{eqnarray}

\begin{figure}[htb]
\includegraphics[width=9cm]{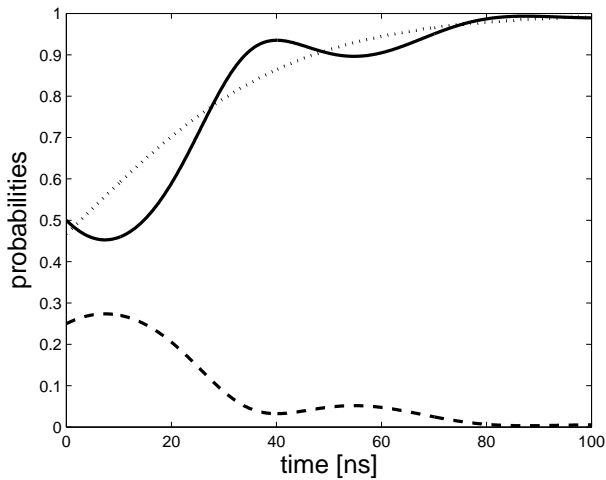}
\caption{The conditional occupation probabilities for the eigenstates $|\lambda_{1,2}'\rangle$ (the dashed line) and $|\lambda_3'\rangle$ (the solid line from $H_\mathrm{cond}$ and the dotted line from Eq. (\ref{eq_entangled_prob})). We have set $E_{J1,2} = \omega_{10} = 10 \ \mathrm{GHz}$, $\gamma_1 = \gamma_2 = 100 \ \mathrm{MHz}$, $\Gamma_1 = 50 \ \mathrm{MHz}$ and $\Gamma = 0.5 \ \mathrm{MHz}$. }
\label{fig_cavity_loss}
\end{figure}

Thus the probability for finding the two qubits in the maximally entangled
state $(|\uparrow\downarrow\rangle - |\downarrow\uparrow\rangle) / \sqrt{2}$, conditioned on that no switching event has been detected, is obtained
from Eq.(\ref{apr}) as
\begin{eqnarray}
P(t) &\approx & \left| \frac{e^{-i\lambda_3' t}}{\sqrt{2}} \right|^2 \left/ \left( \left| \frac{i\gamma e^{-i\lambda_1' t}}{\sqrt{4\gamma^2 - 2\Gamma_1^2}} \right|^2 \right. \right. \nonumber \\
&& \ \ \ \ \left. + \left|  \frac{i\gamma e^{-i\lambda_2' t}}{\sqrt{4\gamma^2 - 2\Gamma_1^2}} \right|^2 + \left| \frac{e^{-i\lambda_3' t}}{\sqrt{2}} \right|^2 \right) . \label{eq_entangled_prob}
\end{eqnarray}
This expression approximately fits the numerical solution calculated by the conditional Hamiltonian in Eq. (\ref{eq_cond_ham}), as
shown in Fig. \ref{fig_cavity_loss}. On a time scale $\Gamma_1^{-1} \ll t \ll \Gamma^{-1}$, the conditional probability of preparing the
maximally entangled two-qubit state approaches unity.

In a real experiment the fidelity of the states prepared by this procedure will not be exactly 1, due to the detector inefficiency and to the
spontaneous decay of the qubits during the time required to reach the asymptotic regime (less than 100 ns, Fig. \ref{fig_cavity_loss}).
Detector efficiencies (visibilities) as large as 0.89 have been obtained
recently for phase qubits \cite{martinisnew}. The contribution of both these processes to the fidelity of the final state in the asymptotic regime
can be calculated \cite{Plenio}. With parameters for superconducting qubits as given in Fig. \ref{fig_cavity_loss}, we estimate that the fidelity of preparing a Bell state
by this procedure will be as high as 90\%.

We conclude this part of the paper by pointing out that this method of producing entangled states is robust with respect to spurious
excitations in the resonator \cite{proc}
and that can be readily applied to existing experimental
architectures such as phase qubit - resonator - phase qubit \cite{mika} without significant changes in the sample design.

\section{Conclusion}
\label{conclude}

We have analyzed the entangling properties of a system consisting of two superconducting qubits coupled to electromagnetic fields,
both in the quantum and classical cases, and in a variety of experimental situations.
Efficient decoupling of the qubits, two-qubit entanglement and high-fidelity quantum gates can be obtained
in this model. We propose also a quantum nondemolition scheme for creating qubit-qubit maximally entangled states by monitoring
the state of the resonator. Our calculations are not dependent on the physical realization of the qubit and the coupling,
therefore most of our results can be adapted to various qubit species and coupling schemes.

\section{Acknowledgments}
Discussions with J. Martinis are gratefully acknowledged.
This work was supported by the Academy of Finland (Acad. Res. Fellowship 00857 and 
projects 7111994 and 7118122) and by Thailand's Commision on Higher Education.

\appendix

\section{Derivation of the Hamiltonian for a long resonator}
\label{resonator_coupling}

In this appendix, we derive the Hamiltonian for the two SCB charge qubits and the CBJJ capacitively coupled to a 1D stripline resonator.
We assume the length of the resonator is $L$, then take $\tilde{c}$ and $\tilde{l}$ as the capacitance and the inductance per unit length.

We start with the classical Lagrangian of the resonator. At a point $0\leq x\leq L$, the local charge density $q_r(x,t)$ and
phase $\varphi_r(x,t)$ satisfy the following relations
\begin{eqnarray}
\frac{\vartheta_0\partial \varphi_r(x,t)}{\partial t} = \frac{q_r(x,t)}{\tilde{c}} , \ \ \ \frac{\partial q_r(x,t)}{\partial t} = \frac{\vartheta_0 \partial^2 \varphi_r(x,t)}{\tilde{l} \partial x^2} , \nonumber
\end{eqnarray}
where $\vartheta_0 \equiv \hbar / 2e$. The Lagrangian of the resonator can be written as
\begin{equation}
{\cal L}_\mathrm{R} = \frac{\vartheta_0^2}{2} \int_0^L dx \left\{ \tilde{c}\left[ \frac{\partial \varphi_r(x,t)}{\partial t} \right]^2 -\frac{1}{\tilde{l}}\left[ \frac{\partial \varphi_r(x,t)}{\partial x} \right]^2 \right\} .
\end{equation}
Because of the boundary conditions
\begin{equation}
\left. \frac{\partial \varphi_r(x,t)}{\partial x} \right|_{x=0} = \left. \frac{\partial \varphi_r(x,t)}{\partial x} \right|_{x=L} = 0 ,
\end{equation}
$\varphi_r(x,t)$ has
the form $\varphi_r(t)\cos(k_n x)$, with $k_n \equiv n\pi / L$. By adopting simplified notations $\varphi_r(t) \rightarrow \varphi_r$ and $\partial \varphi_r(t) / \partial t \rightarrow \dot{\varphi}_r$, the Lagrangian can be rewritten as
\begin{eqnarray}
{\cal L}_\mathrm{R} &=& \vartheta_0^2 \int_0^L dx \left[ \frac{\tilde{c} \dot{\varphi}_r^2 \cos^2(k_nx)}{2} - \frac{\varphi_r^2 k_n^2 \sin^2(k_nx)}{2\tilde{l}} \right] \nonumber \\
&=& \frac{1}{2}\left( \frac{L\tilde{c}}{2} \right) (\vartheta_0\dot{\varphi}_r)^2 - \frac{1}{2}\left( \frac{n^2\pi^2}{2L\tilde{l}} \right)(\vartheta_0\varphi_r)^2 . \label{eq_lag_resonator}
\end{eqnarray}

The Lagrangians of the SCBs and the CBJJ are as follows,
\begin{eqnarray}
{\cal L}_\mathrm{SCB} &=& \sum_{j=1,2} \frac{C_j}{2}(\vartheta_0 \dot{\varphi}_j)^2 + \sum_{j=1,2} \frac{C_{gj}}{2}(\vartheta_0 \dot{\varphi}_j + V_{gj})^2 \nonumber \\
&& + \sum_{j=1,2} E_{Jj}\cos\varphi_j , \label{eq_lag_scb} \\
{\cal L}_\mathrm{CBJJ} &=& \frac{C}{2}(\vartheta_0 \dot{\varphi}_3)^2 + E_J\cos\varphi_3 + I_b\vartheta_0\varphi_3 , \label{eq_lag_cbjj}
\end{eqnarray}
with the gate voltages $V_{gj} = V_{\mathrm{dc}j} + V_{\mathrm{ac}j}$.

By considering that the SCBs are close to the ends of the resonator $x_1\rightarrow 0, \ x_2\rightarrow L$, the CBJJ coupled to the end of the resonator $x_3 = L$, and taking the mode of the resonator $n = 2$, the Lagrangian for the couplings reads
\begin{eqnarray}
{\cal L}_\mathrm{coup} &=& \sum_{j=1,2,3} \frac{C_{mj}}{2} [\vartheta_0\dot{\varphi}_r(x_j) + \vartheta_0\dot{\varphi}_j]^2 \nonumber \\
&\approx& \sum_{j=1,2,3} \frac{C_{mj}}{2} (\vartheta_0\dot{\varphi}_r + \vartheta_0\dot{\varphi}_j)^2 . \label{eq_lag_coup}
\end{eqnarray}
The total Lagrangian is then
\begin{equation}
{\cal L} = {\cal L}_\mathrm{R} + {\cal L}_\mathrm{SCB} + {\cal L}_\mathrm{CBJJ} + {\cal L}_\mathrm{coup} .
\end{equation}

The conjugate charges are calculated by the Legendre transformation $Q_j = -2en_j = \partial {\cal L} / \partial (\vartheta_0 \dot{\varphi}_j)$ ($j = r,1,2,3$), and the total Hamiltonian
\begin{widetext}
\begin{eqnarray}
H &=& \sum_{j = r,1,2,3} Q_j\vartheta_0\dot{\varphi}_j - {\cal L} \nonumber \\
&=& \frac{(2e)^2}{2}\left( \frac{C_{\Sigma 1}C_{\Sigma 2}C_{\Sigma 3}}{\Pi} \right) n_r^2 + \frac{\hbar^2}{2(2e)^2}\left( \frac{2\pi^2}{L\tilde{l}} \right)\varphi_r^2 + \frac{(2e)^2}{2C_{\Sigma 1}}\left( 1 + \frac{C_{m1}^2C_{\Sigma 2}C_{\Sigma 3}}{\Pi} \right) (n_1 - n_{g1})^2 - E_{J1}\cos\varphi_1 \nonumber \\
&& + \frac{(2e)^2}{2C_{\Sigma 2}} \left( 1 + \frac{C_{m2}^2C_{\Sigma 1}C_{\Sigma 3}}{\Pi} \right) (n_2 - n_{g2})^2 - E_{J2}\cos\varphi_2 + \frac{(2e)^2}{2C_{\Sigma 3}}\left( 1 + \frac{C_{m3}^2C_{\Sigma 1}C_{\Sigma 2}}{\Pi} \right)n_3^2 - E_J\cos\varphi_3 - \frac{\hbar}{2e}I_b \varphi_3 \nonumber \\
&& - \frac{(2e)^2C_{m1}C_{\Sigma 2}C_{\Sigma 3}}{\Pi} n_r(n_1 - n_{g1}) - \frac{(2e)^2C_{m2}C_{\Sigma 1}C_{\Sigma 3}}{\Pi} n_r(n_2 - n_{g2}) - \frac{(2e)^2C_{m3}C_{\Sigma 1}C_{\Sigma 2}}{\Pi} n_r n_3 \nonumber \\
&& + \frac{(2e)^2C_{m1}C_{m3}C_{\Sigma 2}}{\Pi} (n_1 - n_{g1})n_3 + \frac{(2e)^2C_{m2}C_{m3}C_{\Sigma 1}}{\Pi} (n_2 - n_{g2})n_3 + \frac{(2e)^2C_{m1}C_{m2}C_{\Sigma 3}}{\Pi} (n_1 - n_{g1})(n_2 - n_{g2}) , \nonumber \\
\label{eq_classic_ham_scb-res-cbjj}
\end{eqnarray}
\end{widetext}
where $n_{gj}\equiv -C_{gj}V_{gj} / 2e$, and
\begin{eqnarray}
\Pi &\equiv& \left( \frac{L\tilde{c}}{2} + C_{\Sigma 4} \right)C_{\Sigma 1}C_{\Sigma 2}C_{\Sigma 3} - C_{m1}^2C_{\Sigma 2}C_{\Sigma 3} \nonumber \\
&& - C_{m2}^2C_{\Sigma 1}C_{\Sigma 3} - C_{m3}^2C_{\Sigma 1}C_{\Sigma 2} , \nonumber
\end{eqnarray}
with
\begin{eqnarray}
&& C_{\Sigma 1} = C_1 + C_{g1} + C_{m1} , \ \ \ C_{\Sigma 2} = C_2 + C_{g2} + C_{m2} , \nonumber \\
&& C_{\Sigma 3} = C + C_{m3} , \ \ \ \ \ \ \ \ \ \ C_{\Sigma 4} = C_{m1} + C_{m2} + C_{m3} . \nonumber
\end{eqnarray}

By assuming that $C_{mj}\ll C_{\Sigma j}$ ($j=1,2,3$), and defining
\begin{eqnarray}
C_{\Sigma r} \equiv \frac{L\tilde{c}}{2} + C_{\Sigma 4} \ \ \ \ \mathrm{and} \ \ \ \ L_{\Sigma r} \equiv \frac{L\tilde{l}}{2\pi^2} \nonumber
\end{eqnarray}
as the effective capacitance and effective inductance of the resonator, the Hamiltonian in Eq. (\ref{eq_classic_ham_scb-res-cbjj}) can be rewritten as
\begin{eqnarray}
H &\approx & \frac{(2e)^2}{2C_{\Sigma r}} n_r^2 + \frac{\hbar^2}{2(2e)^2L_{\Sigma r}} \varphi_r^2 \nonumber \\
&& + \sum_{j=1,2}\left[ \frac{(2e)^2}{2C_{\Sigma j}}(n_j - n_{gj})^2 - E_{Jj}\cos\varphi_j \right] \nonumber \\
&& + \frac{(2e)^2}{2C_{\Sigma 3}}n_3^2 - E_J\cos\varphi_3 - \frac{\hbar}{2e}I_b\varphi_3 \nonumber \\
&& - \frac{(2e)^2}{C_{\Sigma r}} \left[ \frac{C_{m3}}{C_{\Sigma 3}}n_r n_3 + \sum_{j=1,2} \frac{C_{mj}}{C_{\Sigma j}}n_r (n_j - n_{gj}) \right] \nonumber \\
&& + \frac{(2e)^2 C_{m1}C_{m2}}{C_{\Sigma r} C_{\Sigma 1}C_{\Sigma 2}} (n_1 - n_{g1}) (n_2 - n_{g2}) \nonumber \\
&& + \frac{(2e)^2 C_{m3}}{C_{\Sigma r} C_{\Sigma 3}} \sum_{j=1,2} \frac{C_{mj}}{C_{\Sigma j}} (n_j - n_{gj})n_3 . \label{eq_classical_ham_scb-res-cbjj2}
\end{eqnarray}
For a relatively long resonator $L\tilde{c}\gg C_{\Sigma 4}$, the direct SCB-SCB and SCB-CBJJ couplings, described by the last
two lines in Eq. (\ref{eq_classical_ham_scb-res-cbjj2}), are negligible. To obtain the quantum Hamiltonian, we replace the
variables $n_j,\ \varphi_j$ by the operators $\hat{n}_j,\ \hat{\varphi}_j$ which obey the commutation relation
\begin{equation}
[\hat{\varphi}_j, \hat{n}_k] = i\delta_{jk} , \ \ \ \ (j,k = r,1,2,3) . \label{eq_commutation}
\end{equation}

The quantized Hamiltonian of the resonator is then
\begin{equation}
H_\mathrm{R} = \frac{(2e)^2}{2C_{\Sigma r}} \hat{n}_r^2 + \frac{\hbar^2}{2(2e)^2L_{\Sigma r}} \hat{\varphi}_r^2 = \hbar\omega_r \left( a^\dag a + \frac{1}{2} \right) ,
\end{equation}
where $a^\dag$ (a) is photon creation (annihilation) operator, and the resonance frequency
\begin{equation}
\omega_r = 1 \left / \sqrt{L_{\Sigma r}C_{\Sigma r}} \right. \approx 2\pi \left / L\sqrt{\tilde{l} \tilde{c}} \right. .
\end{equation}
The dimensionless charge operator
\begin{equation}
\hat{n}_r = \frac{i}{2e} \sqrt{\frac{C_{\Sigma r} \hbar \omega_r}{2}}(a^\dag - a) . \label{eq_charge_qr}
\end{equation}

By projecting the Hamiltonian of SCBs onto the charge states $|n_j\rangle$, we can get
\begin{eqnarray}
H_\mathrm{SCB} &=& \sum_{j=1,2} \left[ \frac{(2e)^2}{2C_{\Sigma j}} (\hat{n}_j - n_{gj})^2 - E_{Jj}\cos\hat{\varphi}_j \right] \nonumber \\
&=& \sum_{j=1,2}\left\{ \sum_{n_j} \left[ \frac{(2e)^2}{2C_{\Sigma j}}(n_j - n_{g_j})^2 |n_j\rangle\langle n_j| \right. \right. \nonumber \\
&& - \left. \left. \frac{E_{Jj}}{2}(|n_j + 1\rangle \langle n_j| + |n_j-1\rangle\langle n_j|) \right] \right\} . \ \ \ \ \ \ \ \label{eq_scb_untruncate_ham}
\end{eqnarray}
We also assume that the two SCBs are in charge regime, $E_{C1,2} \equiv 2e^2 / C_{\Sigma 1,2} \gg E_{J1,2}$, so that we truncate (\ref{eq_scb_untruncate_ham}) to the two lowest charge states of each SCB, and obtain the Hamiltonian
\begin{equation}
H_\mathrm{SCB} = \sum_{j=1,2} \left[ E_{Cj}\left( n_{gj} - \frac{1}{2} \right) \sigma_j^z - \frac{E_{Jj}}{2} \sigma_j^x \right] . \ \ \ \label{eq_scb_ham_0}
\end{equation}

Since the DC gate voltages are biased at the charge co-degeneracy point (\ref{eq_charge_degeneracy}) and the AC voltages have the form in Eq. (\ref{eq_ac_voltage}), by transforming the Hamiltonian (\ref{eq_scb_ham_0}) into the uncoupled eigenbasis $\{|\uparrow\uparrow\rangle, |\uparrow\downarrow\rangle, |\downarrow\uparrow\rangle, |\downarrow\downarrow\rangle \}$, we arrive at
\begin{equation}
H_\mathrm{SCB} = \sum_{j=1,2} \left[ E_{Cj}w_j(t) \cos(\omega_dt + \phi_j) \sigma_j^x - \frac{E_{Jj}}{2} \sigma_j^z \right] , \label{eq_scb_ham}
\end{equation}
where $w_{j}(t) \equiv -C_{gj}V_{\mathrm{\mu w}j}(t) / 2e$.

For the CBJJ, we consider $E_C \equiv 2e^2 / C_{\Sigma 3} \ll E_J$. It is better to discuss it in the \lq\lq position\rq\rq space, where $\hat{\varphi}_3$ acts as a position operator. In the bottom of one of its local minima (see Fig. \ref{fig_washboard}), the tilted cosine potential is approximated by a harmonic potential. So the Hamiltonian is approximately
\begin{eqnarray}
H_\mathrm{CBJJ} &=& \frac{(2e)^2}{2C_{\Sigma 3}} \hat{n}_3^2 - E_J\cos\varphi_3 - \frac{\hbar}{2e} I_b\hat{\varphi}_3 \nonumber \\
&=& \frac{\hbar\omega_{10}}{2}(|1\rangle\langle 1| - |0\rangle\langle 0|) = -\frac{\hbar\omega_{10}}{2} \sigma_3^z . \ \ \ \ \label{eq_ham_cbjj}
\end{eqnarray}
Here $|0\rangle$ and $|1\rangle$ indicate the ground and the first excited states of the CBJJ, not charge states anymore. $\hbar\omega_{10}$ is the energy difference between the two states. The dimensionless charge is analogous to the momentum
\begin{equation}
\hat{n}_3 = i\sqrt{\frac{\hbar\omega_{10} C_{\Sigma 3}}{2(2e)^2}} (\sigma_3^+ - \sigma_3^-) = \frac{1}{2e}\sqrt{\frac{\hbar\omega_{10} C_{\Sigma 3}}{2}} \sigma_3^y , \label{eq_charge_n3}
\end{equation}
with the raising operator $\sigma_3^+ = |1\rangle\langle 0|$ and the lowering operator $\sigma_3^- = |0\rangle\langle 1|$.

With these charge operators, we obtain the quantized Hamiltonians for resonator-CBJJ and resonator-SCB couplings from the fourth line in Eq. (\ref{eq_classical_ham_scb-res-cbjj2})
\begin{eqnarray}
H_\mathrm{R-C} &=& \kappa (\sigma_3^+ - \sigma_3^-) (a^\dag - a) , \label{eq_ham_r-c} \\
H_\mathrm{R-S} &=& i \sum_{j=1,2} g_j \left[ \sigma_j^x + 2w_j(t)\cos(\omega_dt + \phi_j) \right] (a^\dag - a) , \nonumber \\
\end{eqnarray}
where the coupling strengths
\begin{equation}
\kappa \approx \frac{\hbar C_{m3}}{2}\sqrt{\frac{2\omega_{10}\omega_r}{C_{\Sigma 3} L\tilde{c}}} , \ \ \ \ g_j \approx e \frac{C_{mj}}{C_{\Sigma j}} \sqrt{\frac{\hbar\omega_r}{L\tilde{c}}} .
\end{equation}

The total Hamiltonian is
\begin{equation}
H = H_\mathrm{R} + H_\mathrm{SCB} + H_\mathrm{CBJJ} + H_\mathrm{R-S} + H_\mathrm{R-C} .
\end{equation}

For a very short resonator $C_{\Sigma r}\rightarrow C_{\Sigma 4}$ and $n_r\rightarrow 0$, the kinetic energy terms in Eq. (\ref{eq_classical_ham_scb-res-cbjj2}) become the same as terms in Eq. (\ref{eq_kinetic_energy}), furthermore, $\hbar\omega_r$ is much larger than the other energies in the system, therefore we can assume that the two SCBs and the CBJJ are directly coupled to each other.

\section{Derivation of the Hamiltonian for a short resonator}
\label{capacitor_coupling}

When the stripline resonator is very short, $L\tilde{c}\ll C_{\Sigma 4}$, the center conductor can be considered as an island.
The SCBs and the CBJJ are capacitively coupled to this island. Instead of deriving the Hamiltonian from the classical Lagrangian of this circuit,
here we use a relatively simpler method.

The system has four nodes. As shown in Fig. \ref{fig_system}, the total charge on node $j$ ($j=1,2,3$) is indicated by $Q_j = -2en_j$ ($2$ for a Cooper pair). The center island acts as the fourth node, and the total charge on it is assumed to be $Q_4 = -2en_4$. Since the total charge on a node is equal to the sum of the charges on all capacitors connected to the node, by denoting the electrostatic potential of node $j$ as $V_j$, we write the charge equations for all the nodes in a matrix form, as
\begin{equation}
\left[ \begin{array}{c}
	Q_1 - C_{g1}V_{g1} \\ Q_2 - C_{g2}V_{g2} \\ Q_3 \\ Q_4
\end{array} \right] = \left[ \begin{array}{cccc}
	C_{\Sigma 1} & 0 & 0 & C_{m1} \\
	0 & C_{\Sigma 2} & 0 & C_{m2} \\
	0 & 0 & C_{\Sigma 3} & C_{m3} \\
	C_{m1} & C_{m2} & C_{m3} & C_{\Sigma 4}
\end{array} \right] \left[ \begin{array}{c}
	V_1 \\ V_2 \\ V_3 \\ V_4
\end{array} \right] . \label{eq_capacitance_matrix}
\end{equation}

The $4\times 4$ matrix in Eq. (\ref{eq_capacitance_matrix}) is called the capacitance matrix $\mathbb{C}$.
The total electrostatic (kinetic) energy of the system can be calculated by using the matrix
\begin{equation}
T = \frac{1}{2} \mathbb{Q}^\mathrm{T} \mathbb{C}^{-1} \mathbb{Q} , \label{eq_electrostatic_energy}
\end{equation}
where $\mathbb{Q}$ denotes the column vector of charges on the left-hand side of Eq. (\ref{eq_capacitance_matrix}). By making the same assumption as in Appendix \ref{resonator_coupling}, $C_{m1,2,3}\ll C_{\Sigma 1,2,3}$, and take $Q_4 = 0$, the result of Eq. (\ref{eq_electrostatic_energy}) is
\begin{eqnarray}
T &=& E_{C1}(n_1 - n_{g1})^2 + E_{C2}(n_2 - n_{g2})^2 + E_C n_3^2 \nonumber \\
&& + [ E_{13} (n_1 - n_{g1}) + E_{23} (n_2 - n_{g2}) ] n_3 \nonumber \\
&& + E_{12}(n_1 - n_{g1})(n_2 - n_{g2}) , \label{eq_kinetic_energy}
\end{eqnarray}
with
\begin{eqnarray}
&& E_{C1} \approx 2e^2/C_{\Sigma 1} , \ \ E_{13} \approx \frac{(2e)^2 C_{m1}C_{m3}}{C_{\Sigma 1}C_{\Sigma 3}C_{\Sigma 4}} , \nonumber \\
&& E_{C2} \approx 2e^2/C_{\Sigma 2} , \ \ E_{23} \approx \frac{(2e)^2 C_{m2}C_{m3}}{C_{\Sigma 2}C_{\Sigma 3}C_{\Sigma 4}} , \nonumber \\
&& E_C \approx 2e^2/C_{\Sigma 3} , \ \ \ E_{12} \approx \frac{(2e)^2 C_{m1}C_{m2}}{C_{\Sigma 1}C_{\Sigma 2}C_{\Sigma 4}} . \label{eq_charging_energies}
\end{eqnarray}

The inductive (potential) energy of the system can be expressed as
\begin{equation}
U = -E_{J1}\cos\varphi_1 - E_{J2}\cos\varphi_2 - E_J\cos\varphi_3 - \frac{\hbar}{2e} I_b\varphi_3 , \label{eq_inductive_energy}
\end{equation}
where $\varphi_j$ ($j=1,2,3$) is the gauge-invariant phase difference across each Josephson junction.

The total classical Hamiltonian is $H = T + U$. To derive the quantum Hamiltonian, we again replace the variables $\varphi_j$ and $n_j$ by operators $\hat{\varphi}_j$ and $\hat{n}_j$.

The quantized Hamiltonians of the SCBs and the CBJJ have the same form as those in Eqs. (\ref{eq_scb_ham}) and (\ref{eq_ham_cbjj}). With the charge operators derived in Appendix \ref{resonator_coupling}, the Hamiltonian for direct coupling between the two SCBs is given by
\begin{eqnarray}
H_\mathrm{S-S} &=& E_{12}(\hat{n}_1 - n_{g1}) (\hat{n}_2 - n_{g2}) \nonumber \\
&=& \frac{E_{12}}{4}\sigma_1^x\sigma_2^x + \frac{E_{12}}{2} w_2(t)\cos(\omega_dt + \phi_2)\sigma_1^x \nonumber \\
&& + \frac{E_{12}}{2} w_1(t)\cos(\omega_dt + \phi_1)\sigma_2^x , \label{eq_s-s_coup}
\end{eqnarray}
and the Hamiltonian for the couplings between the two SCBs and the CBJJ is derived from the second line in Eq. (\ref{eq_kinetic_energy})
\begin{equation}
H_\mathrm{S-C} = -\sum_{j=1,2} \gamma_j \left[ \frac{1}{2}\sigma_j^x\sigma_3^y + w_j(t)\cos(\omega_dt + \phi_j)\sigma_3^y \right] ,
\label{eq_scb_cbjj_coupling}
\end{equation}
where
\begin{equation}
\gamma_j = 2e\frac{C_{m3}}{C_{\Sigma 4}} \left( \frac{C_{mj}}{C_{\Sigma j}} \right) \sqrt{\frac{\hbar\omega_{10}}{2C_{\Sigma 3}}} . \label{eq_scb_cbjj_coupling_strength}
\end{equation}

The total Hamiltonian of this system is finally
\begin{equation}
H = H_\mathrm{SCB} + H_\mathrm{CBJJ} + H_\mathrm{S-S} + H_\mathrm{S-C} .
\end{equation}

\section{Derivation of the effective Hamiltonian in the dispersive regime}
\label{eff_resonator_coupling}

We assume that the Rabi frequencies $\Omega_j(t)$ change adiabatically with respect to the qubit (Larmor) frequency,
so in the following calculations we take $\partial \Omega_j / \partial t = 0$. We use the Baker-Hausdorff formula to expand the transformation to second order
\begin{eqnarray}
H_\mathrm{A} &\approx& H + [H,A] + \frac{1}{2}[[H,A],A] - i\frac{\partial A}{\partial t} \nonumber \\
&\approx& - \sum_{j=1,2} \left[ \frac{E_{Jj}}{2} - \frac{g_j^2}{\Delta_j} (a^\dag a + \frac{1}{2}) \right] \nonumber \sigma_j^z \\
&& + \sum_{j=1,2} \frac{g_j\omega_d}{\Delta_j \omega_r} \Omega_j \sin(\omega_dt + \phi_j) (a^\dag + a) \sigma_j^z \nonumber \\
&& + \sum_{j=1,2}\Omega_j\cos(\omega_d + \phi_j)\sigma_j^x + \omega_r a^\dag a \nonumber \\
&& - \frac{g_1g_2}{2}\left( \frac{1}{\Delta_1} + \frac{1}{\Delta_2} \right) (\sigma_1^+\sigma_2^- + \sigma_1^-\sigma_2^+) .
\end{eqnarray}

By making an adiabatic approximation for the resonator, $\langle a^\dag a\rangle = 0$ and $\langle a^\dag + a\rangle = 1$, we neglect the ac-Stark shift and Lamb shift terms and the Hamiltonian of the resonator. The effective Hamiltonian can be written as
\begin{eqnarray}
H_A = H_{\mathrm{A}0} + H_{\mathrm{A}1} , \nonumber
\end{eqnarray}
where
\begin{eqnarray}
H_{A0} = \sum_{j=1,2}\frac{g_j\Omega_j \omega_d}{\Delta_j\omega_r}\sin(\omega_dt + \phi_j) \sigma_j^z ,
\end{eqnarray}
and
\begin{eqnarray}
H_{A1} &=& \sum_{j = 1,2}\left[ -\frac{E_{Jj}}{2}\sigma_j^z + \Omega_j\cos(\omega_dt + \phi_j)\sigma_j^x \right] \nonumber \\
&& - \frac{g_1g_2(\Delta_1 + \Delta_2)}{2\Delta_1 \Delta_2}(\sigma_1^+\sigma_2^- + \sigma_1^-\sigma_2^+).
\end{eqnarray}

Now we transform $H_{A1}$ into the interaction picture
\begin{eqnarray}
\widetilde{H} = e^{i\int_0^t dt' H_{A0}(t')}H_{A1}e^{-i\int_0^t dt' H_{A0}(t')} .
\end{eqnarray}
The raising and lowering operators become
\begin{eqnarray}
\tilde{\sigma}_j^\pm &=& \sigma_j^\pm e^{\pm i z\cos(\omega_dt + \phi_j)} \nonumber \\
&=& \sum_{n=-\infty}^\infty i^n J_n(\pm z) e^{in(\omega_dt + \phi_j)} ,
\end{eqnarray}
where $J_n(\pm z)$ is the $n$-th order Bessel function of the first kind. Since $z = g_j\Omega_j / (\Delta_j\omega_r) \ll 1$, $J_0(\pm z)$ is dominant. Therefore $\tilde{\sigma}_j^\pm \approx \sigma_j^\pm$, and the Hamiltonian
\begin{eqnarray}
\widetilde{H} &\approx& \sum_{j=1,2}\left[ -\frac{E_{Jj}}{2}\sigma_j^z + \Omega_j\cos(\omega_dt + \phi_j)\sigma_j^x \right] \nonumber \\
&& - \frac{g_1g_2(\Delta_1 + \Delta_2)}{2\Delta_1 \Delta_2}(\sigma_1^+\sigma_2^- + \sigma_1^-\sigma_2^+) .
\end{eqnarray}

\section{Leakage out of the computational subspace}
\label{leakage}

Here we only consider a single charge qubit with rectangular driving pulse $\Omega \cos(\omega_d t)$. By considering the three lowest charge states $|n\rangle = \{ |0\rangle, |1\rangle, |2\rangle \}$, and dc biasing to the degeneracy point, the Hamiltonian has the form
\begin{eqnarray}
H &=& E_C(\hat{n}-n_g)^2 - E_J\cos\hat{\varphi} \nonumber \\
&=& \left[
 	\begin{array}{ccc}
        0 & -E_J/2 & 0 \\
        -E_J/2 & -2\Omega\cos(\omega_d t) & -E_J/2 \\
        0 & -E_J/2 & 2E_C - 4\Omega\cos(\omega_d t)
        \end{array}
    \right] . \nonumber
\end{eqnarray}

\begin{figure}[htb]
\includegraphics[width=9cm]{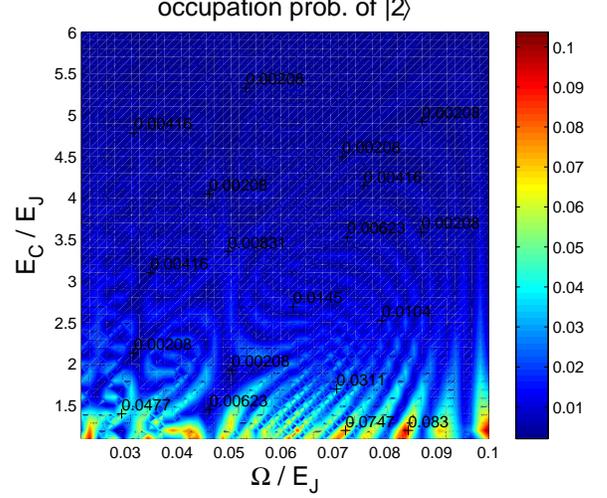}
\caption{(Color online) Leakage for a $\pi$-pulse.} \label{fig_leakage}
\end{figure}

The leakage can be defined as the occupation probability of state $|2\rangle$ after certain pulse duration, averaged overall possible initial states $|\psi_\mathrm{in} \rangle = \cos\frac{\theta}{2}|0\rangle + e^{i\phi}\sin\frac{\theta}{2}|1\rangle$. In Fig. \ref{fig_leakage}, the leakage for a $\pi$-pulse is plotted by numerically solving the Schr\"odinger equation with $H$. The driving frequency is set to $\omega_d = E_J$.

\section{Qubit-TLS system}
\label{qubit_tls}

A lot of experimental progress has been made recently on phase qubits following the realization that the
dielectric insulator forming the Josephson junction contains two-level system (TLS) defects \cite{phase,junctionresonators}.
These defects have been shown to have decoherence times comparable to that of the qubit, thus they can be addressed coherently ({\it e.g.} by tuning the qubit on- and off- resonance with them).

The form of the interaction Hamiltonian between the qubit and the TLS is of the type
$\sigma_{x}\sigma_{x}$ in the case of phase qubits \cite{junctionresonators}, and this coupling becomes important
when $\Delta \equiv |\omega_1 - \omega_2|\lesssim \omega^{xx}$. Here we adopt the same notations as in Sec. \ref{switchable},
 $\omega^{xx}$ denotes the coupling strength between the qubit and the TLS, $\omega_1$ and $\omega_2$ are Larmor frequencies of the
 qubit and the TLS, respectively. By assuming that for a single qubit there is only one such TLS near it, and the TLS is
 weakly coupled to the driving field of the qubit, we may use the Hamiltonian in Eq. (\ref{eq_qubit_tls_resonator}) to describe this
 qubit-TLS system, and therefore use the switchable scheme developed in Sec. \ref{high_osc_freq} to perform quantum gates
 with the qubit and the TLS.


\begin{thebibliography}{99}
\bibitem{Cartan}
$U_\mathrm{CNOT} = U_1 M U_2$, where $U_{1,2}\in \mathrm{SU}(2)\otimes \mathrm{SU}(2)$ are two single-qubit gates, and $M$ is called the locally equivalent gate of CNOT. See Refs. \cite{Makhlin, Vidal}.
\bibitem{note2}
We assume that the temperature is lower than $\hbar\omega_{10} / 2\pi k_B$, so that the thermal activation is neglected.
\end{thebibliography}
\end{document}